\input harvmac.tex

\input epsf

\def\figin{\epsfcheck\figin}\def\figins{\epsfcheck\figins}
\def\epsfcheck{\ifx\epsfbox\UnDeFiNeD
\message{(NO epsf.tex, FIGURES WILL BE IGNORED)}
\gdef\figin##1{\vskip2in}\gdef\figins##1{\hskip.5in}
\else\message{(FIGURES WILL BE INCLUDED)}%
\gdef\figin##1{##1}\gdef\figins##1{##1}\fi}
\def\DefWarn#1{}
\def\figinsert{\goodbreak\midinsert}
\def\ifig#1#2#3{\DefWarn#1\xdef#1{fig.~\the\figno}
\writedef{#1\leftbracket fig.\noexpand~\the\figno}%
\figinsert\figin{\centerline{#3}}\medskip\centerline{\vbox{\baselineskip12pt
\advance\hsize by -1truein\noindent\footnotefont{\bf Fig.~\the\figno:} #2}}
\bigskip\endinsert\global\advance\figno by1}

\noblackbox
%


\def\unlockat{\catcode`\@=11}
\def\lockat{\catcode`\@=12}

\unlockat

\def\newsec#1{\global\advance\secno by1\message{(\the\secno. #1)}
\global\subsecno=0\global\subsubsecno=0\eqnres@t\noindent
{\bf\the\secno. #1}
\writetoca{{\secsym} {#1}}\par\nobreak\medskip\nobreak}
\global\newcount\subsecno \global\subsecno=0
\def\subsec#1{\global\advance\subsecno
by1\message{(\secsym\the\subsecno. #1)}
\ifnum\lastpenalty>9000\else\bigbreak\fi\global\subsubsecno=0
\noindent{\it\secsym\the\subsecno. #1}
\writetoca{\string\quad {\secsym\the\subsecno.} {#1}}
\par\nobreak\medskip\nobreak}
\global\newcount\subsubsecno \global\subsubsecno=0
\def\subsubsec#1{\global\advance\subsubsecno by1
\message{(\secsym\the\subsecno.\the\subsubsecno. #1)}
\ifnum\lastpenalty>9000\else\bigbreak\fi
\noindent\quad{\secsym\the\subsecno.\the\subsubsecno.}{#1}
\writetoca{\string\qquad{\secsym\the\subsecno.\the\subsubsecno.}{#1}}
\par\nobreak\medskip\nobreak}

\def\subsubseclab#1{\DefWarn#1\xdef
#1{\noexpand\hyperref{}{subsubsection}%
{\secsym\the\subsecno.\the\subsubsecno}%
{\secsym\the\subsecno.\the\subsubsecno}}%
\writedef{#1\leftbracket#1}\wrlabeL{#1=#1}}
\lockat


\def\boxit#1{\vbox{\hrule\hbox{\vrule\kern8pt
\vbox{\hbox{\kern8pt}\hbox{\vbox{#1}}\hbox{\kern8pt}}
\kern8pt\vrule}\hrule}}
\def\mathboxit#1{\vbox{\hrule\hbox{\vrule\kern8pt\vbox{\kern8pt
\hbox{$\displaystyle #1$}\kern8pt}\kern8pt\vrule}\hrule}}
%
\def\exercise#1{\bgroup\narrower\footnotefont
\baselineskip\footskip\bigbreak
\hrule\medskip\nobreak\noindent {\bf Exercise}. {\it #1\/}\par\nobreak}
\def\endexercise{\medskip\nobreak\hrule\bigbreak\egroup}
%

%
%
%
\def\CC{{\cal C}}

\def\CH{{\cal H}}

\def\CM{{\cal M}}
\def\CN{{\cal N}}

\def\IZ{\relax\ifmmode\mathchoice
{\hbox{\cmss Z\kern-.4em Z}}{\hbox{\cmss Z\kern-.4em Z}}
{\lower.9pt\hbox{\cmsss Z\kern-.4em Z}}
{\lower1.2pt\hbox{\cmsss Z\kern-.4em Z}}\else{\cmss Z\kern-.4em
Z}\fi}
\def\inbar{\,\vrule height1.5ex width.4pt depth0pt}
\def\IB{\relax{\rm I\kern-.18em B}}
\def\IC{\relax\hbox{$\inbar\kern-.3em{\rm C}$}}
\def\ID{\relax{\rm I\kern-.18em D}}
\def\IE{\relax{\rm I\kern-.18em E}}
\def\IF{\relax{\rm I\kern-.18em F}}
\def\IG{\relax\hbox{$\inbar\kern-.3em{\rm G}$}}
\def\IH{\relax{\rm I\kern-.18em H}}
\def\II{\relax{\rm I\kern-.18em I}}
\def\IK{\relax{\rm I\kern-.18em K}}
\def\IL{\relax{\rm I\kern-.18em L}}
\def\IM{\relax{\rm I\kern-.18em M}}
\def\IN{\relax{\rm I\kern-.18em N}}
\def\IO{\relax\hbox{$\inbar\kern-.3em{\rm O}$}}
\def\IP{\relax{\rm I\kern-.18em P}}
\def\IQ{\relax\hbox{$\inbar\kern-.3em{\rm Q}$}}
\def\IR{\relax{\rm I\kern-.18em R}}
\font\cmss=cmss10 \font\cmsss=cmss10 at 7pt
\def\IZ{\relax\ifmmode\mathchoice
{\hbox{\cmss Z\kern-.4em Z}}{\hbox{\cmss Z\kern-.4em Z}}
{\lower.9pt\hbox{\cmsss Z\kern-.4em Z}}
{\lower1.2pt\hbox{\cmsss Z\kern-.4em Z}}\else{\cmss Z\kern-.4em Z}\fi}
\def\IGa{\relax\hbox{${\rm I}\kern-.18em\Gamma$}}
\def\IPi{\relax\hbox{${\rm I}\kern-.18em\Pi$}}
\def\ITh{\relax\hbox{$\inbar\kern-.3em\Theta$}}
\def\IOm{\relax\hbox{$\inbar\kern-3.00pt\Omega$}}

\def\inbar{\,\vrule height1.5ex width.4pt depth0pt}

\font\cmss=cmss10 \font\cmsss=cmss10 at 7pt
\def\IR{\relax{\rm I\kern-.18em R}}

\def\Tr{\rm Tr}
\def\vol{{\rm vol}}

\def\im{{\rm Im}}

\def\gsim{
{\ \lower-1.2pt\vbox{\hbox{\rlap{$>$}\lower5pt\vbox{\hbox{$\sim$}}}}\ }
}
\def\lsim{{\
\lower-1.2pt\vbox{\hbox{\rlap{$<$}\lower5pt\vbox{\hbox{$\sim$}}}}\ }
}

\lref\bcov{M. Bershadsky, S. Cecotti, H.Ooguri and C. Vafa, ``Holomorphic
anomalies in
topological field theory," Nucl. Phys. {\bf B 405} (1993) 279.}

\lref\lick{W.B.R. Lickorish, {\it An introduction to knot theory},
Springer, 1997.}

\lref\gottsche{L. G\"ottsche, ``Modular forms and Donaldson
invariants for 4-manifolds with $b_+=1$,'' alg-geom/9506018; J. Am. Math. Soc.
{\bf 9}
(1996) 827.}

\lref\gottzag{L. G\"ottsche and D. Zagier,
``Jacobi forms and the structure of Donaldson
invariants for 4-manifolds with $b_+=1$,''
alg-geom/9612020.}

\lref\mw{G. Moore and E. Witten, ``Integration over
the $u$-plane in Donaldson theory," hep-th/9709193.}

\lref\swi{N. Seiberg and E. Witten,
``Electric-magnetic duality, monopole condensation, and confinement in
${\cal N}=2$ supersymmetric Yang-Mills
Theory,''
hep-th/9407087; Nucl. Phys. {\bf B426} (1994) 19}

\lref\swii{N. Seiberg and E. Witten,
``Monopoles, Duality and Chiral Symmetry Breaking in N=2 Supersymmetric QCD,''
hep-th/9408099}

\lref\vw{C. Vafa and E. Witten,
``A strong coupling test of $S$-duality,''
hep-th/9408074; Nucl. Phys. {\bf B431} (1994) 3.}

\lref\monopole{E. Witten, ``Monopoles and
four-manifolds,''  hep-th/9411102; Math. Res. Letters {\bf 1} (1994) 769.}

\lref\witteni{E. Witten, ``On $S$-duality in abelian
gauge theory,'' hep-th/9505186; Selecta Mathematica {\bf 1} (1995) 383.}

\lref\wittk{E. Witten, ``Supersymmetric Yang-Mills theory
on a four-manifold,''  hep-th/9403193;
J. Math. Phys. {\bf 35} (1994) 5101.}

\lref\zagi{D. Zagier, ``Nombres de classes et formes
modulaires de poids 3/2,'' C.R. Acad. Sc. Paris,
{\bf 281A} (1975)883.}
\lref\zagii{F. Hirzebruch and D. Zagier,
``Intersection numbers of curves on Hilbert modular
surfaces and modular forms of Nebentypus,''
Inv. Math. {\bf 36}(1976)57.}

\lref\lns{A. Losev, N. Nekrasov, and S. Shatashvili, ``Issues in
topological gauge theory," hep-th/9711108; ``Testing Seiberg-Witten solution,"
hep-th/9801061.}

\lref\mmone{M. Mari\~no and G. Moore, ``Integrating over the Coulomb branch in
${\cal N}=2$ gauge theory," hep-th/9712062.}

\lref\mmtwo{M. Mari\~no and G. Moore, ``The Donaldson-Witten function for gauge
groups
of rank larger than one," hep-th/9802185.}

\lref\munnsc{V. Mu\~noz, ``Basic classes for four-manifolds
not of simple type,'' math.DG/9811089.}

\lref\fandm{R. Friedman and J.W. Morgan,
``Algebraic surfaces and Seiberg-Witten invariants,''
alg-geom/9502026; J. Alg. Geom. {\bf 6} (1997) 445. ``Obstruction bundles,
semiregularity, and Seiberg-Witten
invariants'', alg-geom/9509007.}

\lref\morganbk{J.W. Morgan, {\it The Seiberg-Witten equations and applications
to the topology of smooth four-manifolds}, Princeton University Press, 1996.}

\lref\DoKro{S.K.~ Donaldson and P.B.~ Kronheimer,
{\it The Geometry of Four-Manifolds},
Clarendon Press, Oxford, 1990.}

\lref\FrMor{R. Friedman and J.W. Morgan,
{\it Smooth Four-Manifolds and Complex Surfaces},
Springer Verlag, 1991.}

\lref\tqft{E. Witten,
``Topological Quantum Field Theory,''
Comm. Math. Phys. {\bf 117} (1988)
353.}

\lref\mt{G. Meng and C. Taubes,  ``${\underline {SW}}$ =Milnor torsion",
Math. Res. Lett.  {\bf 3} (1996) 661. }

\lref\gorsky{A. Gorsky, A. Marshakov, A. Mironov and A.Morozov, ``RG equations
from
Whitham hierarchy," hep-th/9802007.}

\lref\turaev{V.G. Turaev, ``Reidemeister torsion in knot theory," Russ.
Math. Surveys {\bf 41} (1986) 119.}

\lref\mst{J. Morgan, T. Szab\'o and C.H. Taubes,
``A product formula for the Seiberg-Witten
invariants and the generalized Thom conjecture,"
J. Diff. Geometry {\bf 44} (1996) 706.}

\lref\ms{J.W. Milnor and J.D. Stasheff, {\it Characteristic classes},
Princeton University Press, 1974.}

\lref\doncobord{S. Donaldson,
``Irrationality and the $h$-cobordism conjecture,''
J. Diff. Geom. {\bf 26} (1987) 141.}

\lref\donaldsonii{S. Donaldson, ``Floer homology and
algebraic geometry,'' in {\it Vector bundles in
algebraic geometry}, N.J. Hitchin et. al. eds.
Cambridge 1995.}

\lref\floer{{\it The Floer Memorial Volume}, H. Hofer et. al.
eds., Birk\"auser 1995.}

\lref\munoz{V. Mu\~noz, ``Wall-crossing formulae for algebraic surfaces with
$q>0$," alg-geom/9709002.}

\lref\munozfloer{V. Mu\~noz, ``Ring structure of the Floer cohomology of
$\Sigma \times {\bf S}^1$," dg-ga/9710029.}

\lref\munozqu{V. Mu\~noz, ``Quantum cohomology of the moduli space of stable
bundles over a Riemann surface," alg-geom/9711030.}

\lref\liliu{T.J. Li and A. Liu, ``General wall-crossing formula,'' Math. Res.
Lett. {\bf 2} (1995) 797.}

\lref\sh{S. Shenker, ``Another length scale in string theory?,"
hep-th/9509132.}

\lref\ssh{N. Seiberg and S. Shenker, ``Hypermultiplet moduli space and
string compactification to three dimensions," hep-th/9608086, Phys. Lett.
{\bf B 388} (1996) 521.}

\lref\okonek{C. Okonek and A. Teleman, ``Seiberg-Witten invariants for
manifolds with $b_2^+=1$, and the universal wall-crossing formula,"
alg-geom/9603003; Int. J. Math. {\bf 7} (1996) 811.}

\lref\carey{A.L. Carey, B.L. Wang, R.B. Zhang and J. McCarthy,
``Seiberg-Witten monopoles in three dimensions," Lett.
Math. Phys. {\bf 39} (1997) 213. Y. Ohta, ``Topological Field
Theories associated with three-dimensional Seiberg-Witten monopoles,"
Int. J. Theor. Phys. {\bf 37} (1998) 925.}

\lref\verlinde{E. Verlinde, ``Global aspects of electric-magnetic duality,"
hep-th/9506011;
Nucl. Phys. {\bf B455} (1995) 211. }

\lref\lescop{C. Lescop, {\it Global surgery formula for the Casson-Walker
invariant}, Annals
of Mathematical Studies, Princeton University Press, 1996.}

\lref\swgr{C. Taubes, ``The Seiberg-Witten and the Gromov invariants,"
Math. Res. Lett. {\bf 2} ; ``${\rm SW}\Rightarrow
{\rm Gr}$: from Seiberg-Witten invariants to holomorphic curves," J. Amer.
Math. Soc. (1996) .}

\lref\gthompson{G. Thompson, ``On the Generalized
Casson Invariant,'' hep-th/9811199}

\lref\threekron{P. Kronheimer, ``Embedded surfaces and gauge theory in
three and four
dimensions", preprint.}

\lref\khoze{
N. Dorey, V. V. Khoze, M. P. Mattis, D. Tong, S. Vandoren,
``Instantons, Three-Dimensional Gauge Theory, and the Atiyah-Hitchin
Manifold,''
hep-th/9703228; Nucl. Phys. {\bf B 502} (1997) 59.}

\lref\marwang{M. Marcolli, ``Seiberg-Witten-Floer
homology and Heegard splittings," Int. J. Math. {\bf 7} (1996)
671. M. Marcolli and B.-L. Wang, ``Equivariant Seiberg-Witten-Floer homology,"
dg-ga/9606003.}

\lref\rw{L. Rozansky and E. Witten, ``Hyperk\"ahler
geometry and invariants of three manifolds,"
hep-th/9612126.}

\lref\threesw{N. Seiberg and E. Witten, ``Gauge dynamics
and compactification to three dimensions," hep-th/9607163.}

\lref\ffloer{V. Mu\~noz, ``Fukaya-Floer homology of
$\Sigma \times {\bf S}^1$ and applications,"
dg-ga/9804081.}

\lref\ns{M. Mari\~no and G. Moore, ``Donaldson invariants for non-simply
connected
manifolds," hep-th/9804104.}

\lref\roz{L. Rozansky, ``A contribution of the trivial connection to the
Jones polynomial and Witten's invariant of 3d manifolds, " Comm. Math.
Phys. {\bf 175} (1996) 275.}

\lref\kots{D. Kotschick, ``$SO(3)$-invariants for manifolds with
$b_2^+=1$," Proc. London Math. Society {\bf 63} (1991) 426.}

\lref\floerb{A. Floer, ``Instanton homology and Dehn surgery," in {\it The
Floer memorial volume},
Birkh\"auser, 1994.}

\lref\topc{E. Witten, ``Topology changing amplitudes in $(2+1)$ dimensional
gravity, " Nucl. Phys. {\bf B 323} (1989) 113. }

\lref\aj{M.F. Atiyah and L. Jeffrey, ``Topological Lagrangians and
cohomology," J. Geom. Phys. {\bf 7} (1990) 119.}

\lref\bt{M. Blau and G. Thompson, ``${\cal N}=2$ topological gauge theory,
the Euler characteristic of moduli spaces, and the Casson invariant," Comm.
Math. Phys. {\bf 152} (1993) 41.}

\lref\tqft{E. Witten,
``Topological Quantum Field Theory,''
Comm. Math. Phys. {\bf 117} (1988)
353.}

\lref\egh{T.~ Eguchi, P.B.~ Gilkey, and A.J.~ Hanson,
``Gravitation, Gauge Theories, and  Differential Geometry",
 Phys. Rep. {\bf 66}(1980) 214.}

\lref\btft{R. Dijkgraaf and G. Moore,
``Balanced Topological Field Theories,''
hep-th/9608169; Comm. Math. Phys. {\bf 185} (1997) 411.
}

\lref\seibtd{N. Seiberg, ``IR Dynamics on Branes and Space-Time Geometry,''
hep-th/9606017; Phys.Lett. B384 (1996) 81-85}

\lref\giveonkutasov{A. Giveon and D. Kutasov,
``Brane Dynamics and Gauge Theory,''  hep-th/9802067}

\lref\wittfdsol{E. Witten, ``Solutions Of Four-Dimensional Field Theories
Via M Theory,''    hep-th/9703166; Nucl. Phys. {\bf B 500} (1997) 3.}

\lref\hanwit{A. Hanany and E. Witten,
``Type IIB Superstrings, BPS Monopoles, And Three-Dimensional Gauge Dynamics,''
hep-th/9611230; Nucl. Phys. {\bf B 492} (1997) 152.}

\lref\jonespoly{E.~ Witten, ``Quantum Field Theory and  the Jones Polynomial",
Comm. Math. Phys. {\bf 121} (1989) 351.}

\Title{\vbox{\baselineskip12pt
\hbox{hep-th/9811214}
\hbox{YCTP-P27-98   }
}}
{\vbox{\centerline{3-manifold topology}
\medskip
\centerline{ and the }
\medskip
\centerline{   Donaldson-Witten partition function}
 }}

\bigskip
\centerline{Marcos Mari\~no and   Gregory Moore}
\medskip
\centerline{Department of Physics, Yale University,}
\centerline{New Haven, CT 06520}
\centerline{ \it marcos.marino@yale.edu }
\centerline{ \it moore@castalia.physics.yale.edu }

\bigskip
We consider Donaldson-Witten theory
on four-manifolds of the form $X=Y \times {\bf S}^1$
where $Y$ is a compact three-manifold.
We show that there are interesting relations
between the four-dimensional
Donaldson invariants of $X$ and
certain topological invariants of $Y$.
In particular, we reinterpret a
result of Meng-Taubes relating the
Seiberg-Witten invariants to Reidemeister-Milnor
torsion. If $b_1(Y)>1$
we show that the partition function reduces
to the  Casson-Walker-Lescop invariant of $Y$,
as expected on formal grounds.
In the case $b_1(Y)=1$ there is a correction.
Consequently, in the case  $b_1(Y)=1$,  we
observe an interesting  subtlety in the
standard expectations of Kaluza-Klein theory
when applied to
supersymmetric gauge theory compactified
on a circle of small radius.

\Date{Nov. 23, 1998}

\newsec{Introduction}

The application of   gauge theory to the
topology of 3- and 4-manifolds has proved to be a
deep and rich subject. This subject is now
mature and has an extensive literature.
 Nevertheless, important unsolved problems
and challenges remain. Many of these
open problems are related to Floer homology, and
the topology of three-manifolds.

Last year,  there was some
progress in the application of Witten's physical
approach to Donaldson theory via supersymmetric
gauge theory \tqft\wittk\monopole. This progress was the
result of an improved understanding
of $\CN=2$ supersymmetric Yang-Mills theory on
4-manifolds $X$ with $b_2^+(X)=1$ \mw.
For example, the extension of Donaldson-Witten
theory to nonsimply connected manifolds was
recently completed in \ns.

In the present paper we
continue to explore some of the consequences of
the results of \ns,  focusing on the applications to
3-manifold topology. In particular,
   we discuss the evaluation of
the Donaldson-Witten function
 $Z_{DW}(Y \times {\bf S}^1)$  for
3-manifolds $Y$ with $b_1(Y)>0$.
The function $Z_{DW}(X)$ for a four-manifold
$X$ is the partition function of a
topologically-twisted $d=4,\CN=2$
gauge theory on $X$. In this paper we
concentrate on   the
gauge groups $SU(2)$ and $SO(3)$,
although generalizations should be
accessible using \mmtwo.

The supersymmetric Yang-Mills
approach is particularly well-suited to addressing
questions related to Floer homology since the
Floer groups are the Hilbert spaces of quantum
ground states of the topological field theory \tqft.
Given a 3-fold $Y$
 the path integral $Z_{DW}(Y \times {\bf S}^1)$ is
just the Witten index  $\Tr (-1)^F$ of the
theory and therefore is the Euler character of
the Floer homology. The arguments for this
are of course formal, and we check the
result for $b_1(Y)>1$ in section 4.2.

In section five we move on to the more difficult case
of $b_1(Y)=1$.  In this case we encounter an
interesting subtlety in the compactification of
supersymmetric Yang-Mills theory on a circle.
This subtlety  is connected with the
noncompact nature  of the moduli space of
SYM groundstates.   We consider a  product
metric on   $X= Y \times {\bf S}^1$ where the ${\bf S}^1$
factor has radius $R$. Taking the $R \rightarrow 0$ limit
one finds an effective 3-dimensional theory
described in detail in \threesw. When topologically
twisted, the partition function of
 this theory defines the
Rozansky-Witten invariant of $Y$ \rw. On
the other hand we may evaluate
$Z_{DW}(Y \times {\bf S}^1)$ directly and take the
$R \rightarrow 0$ limit. For 3-folds $Y$ with
$b_1(Y)=1$ the answer does not agree with
the Rozansky-Witten invariant. Thus,
{\it the naive Kaluza-Klein expectation fails
for this class of observables}. We discuss some
of the physics associated with this surprising
fact in section 6. Our main conclusion is that
the theory must be simultaneously regarded as
three- and four-dimensional.

\newsec{A brief review of some three-manifold invariants}

We will need some basic facts and results about  Reidemeister-Milnor
torsion and
Alexander polynomials of   three manifolds as well as some
related results on  knot and link invariants. We review these
for the reader's convenience. Good references for these results are
\turaev\lescop\lick.

Let $Y$ be a connected,  oriented three-dimensional manifold with
$b_1(Y)>0$. The situations we will have in mind are compact manifolds or
the complement of a link in a compact manifold. The invariants of $Y$
depend in part on
the structure of the first homology group of
$Y$, $H_1 (Y, \IZ)$. This
is a finitely generated abelian group. Accordingly
it has a free part and a torsion part:
\eqn\freetor{
H_1 (Y, \IZ) = \IZ^{b_1(Y)} \oplus {\rm Tor}(H_1(Y, \IZ))
}
where ${\rm Tor}(H_1(Y, \IZ))$ is the subgroup of
elements of finite order.
We denote the associated
 free abelian group of rank $b_1(Y)$  by
\eqn\defineh{
H(Y)\equiv H_1(Y, \IZ)/{\rm Tor}(H_1(Y, \IZ))\cong \IZ^{b_1(Y)}.
}
Note that, by the universal coefficient theorem, $H(Y) \simeq H^2(Y,
\IZ)/{\rm Tor} (H^2(Y, \IZ))$.
Here and below we will sometimes simply
write $H$ when no confusion can arise.
We will also denote the generators of $H$ by $h_1, \dots, h_{b_1}$. By
Poincar\'e duality, this gives a basis for the free part of $H^2(Y, \IZ)$
that will be denoted by $h_i$ as well,
with $i=1, \dots, b_1\equiv b_1(Y)$.

The invariants of $Y$ that we will discuss
 are multivariable
Laurent polynomials and Laurent series in
variables $t_i$ which can be formally identified as
 $t_i= \exp (h_i)$. (We are being somewhat imprecise
here for the sake of expediency. Fussbudgets
should consult the remark at the end of this section.)
The Reidemeister-Milnor torsion of $Y$,
which we will denote by
$\tau(Y;t_i)$, is in general  a formal power series in $t_i$, $t_i^{-1}$.
The precise definition of this invariant, (which we will not need)
 can be found in \turaev. Roughly speaking, one
introduces a combinatorial description of $Y$ and considers
ratios of volumes of the terms in the associated
(acyclic) chain complexes.

One of the basic properties of the torsion is
that
it is symmetric under the exchange $t_i \leftrightarrow t_i^{-1}$. The
Reidemeister-Milnor torsion of $Y$ is related to the analytic Ray-Singer
torsion, which is defined in terms of the Laplacian operator on forms with
coefficient in a flat bundle. For instance, if $b_1 (Y)=1$, and we consider
the flat $U(1)$ connection $A_{\mu}= u \omega_{\mu}$, where $\omega$ is a
harmonic one form that represents a generator of $H^1 (Y, \IR)$, then
$|\tau(Y; {\rm e}^{iu})|$ is the Ray-Singer torsion associated to the flat
connection. Similar considerations apply when $b_1(Y)>1$. This equivalence
is important in the analysis of perturbative Chern-Simons theory
\jonespoly\roz.

Another classical invariant of three manifolds is the Alexander polynomial
of $Y$, denoted by $\Delta_Y(t_i)$, which is a
Laurent {\it polynomial} in the $t_i$.  The Reidemeister-Milnor torsion is
related to the Alexander polynomial as follows (see \turaev, Theorem 1.1.2):
if $b_1(Y)>1$, one has
\eqn\torgo{
\tau (Y; t_i) = \Delta_Y (t_i), \,\,\,\,\,\, b_1(Y)>1.}
In particular, for $b_1 (Y)>1$,
the Reidemeister-Milnor torsion is   a
Laurent polynomial.

On the other hand, if $b_1(Y)=1$, then
\eqn\torso{
\tau(Y;t) = {\Delta_Y (t) \over (t^{1/2}-t^{-1/2})^{\epsilon (Y)}},
\,\,\,\,\, b_1(Y)=1 ,}
where $\epsilon (Y) =2$ if $Y$ is closed and $1$ if $Y$ has a boundary.
\foot{The definition of the torsion for manifolds with boundary requires
some discussion of boundary conditions that we don't need here.}
Note that for $b_1 (Y)=1$ the torsion $\tau(Y)$ is an
infinite Laurent series.
For $b_1(Y)=1$ the Alexander polynomial is  also   symmetric   under $t
\rightarrow t^{-1}$ with the structure
\eqn\alexander{\Delta_Y(t)= a_0 + \sum_{i=1}^r a_i (t^i+t^{-i}).
}
Moreover, it satisfies
 the important property (\turaev, Theorem 1.6.1)
\eqn\torsum{
\Delta_Y(1)=|{\rm Tor}(H_1(Y, \IZ))|.}
Using this
result, the Reidemeister-Milnor
torsion for a closed manifold $Y$ with $b_1(Y)=1$ can be rewritten in the form:
\eqn\milnor{
\tau(Y;t)= { |{\rm Tor}(H_1(Y, \IZ))| \over  (t^{1/2} -t^{-1/2})^2}
+\sum_{j=1}^r j a_j+  \sum_{k=1}^{r-1} \biggl[
\bigl( \sum_{j=1}^{r-k} j a_{j+k}\bigr)  (t^k + t^{-k}) \biggr].
}
This representation will prove useful in some of the
derivations below.

\ifig\meridian{The thin dashed line is the component of
a link, surrounded by a solid torus. The two thick dashed
lines are on the boundary of the torus. The longitude
runs parallel to the link-component (``b-cycle'') and the meridian
runs horizontal (``a-cycle''). }
{\epsfxsize2.5in\epsfbox{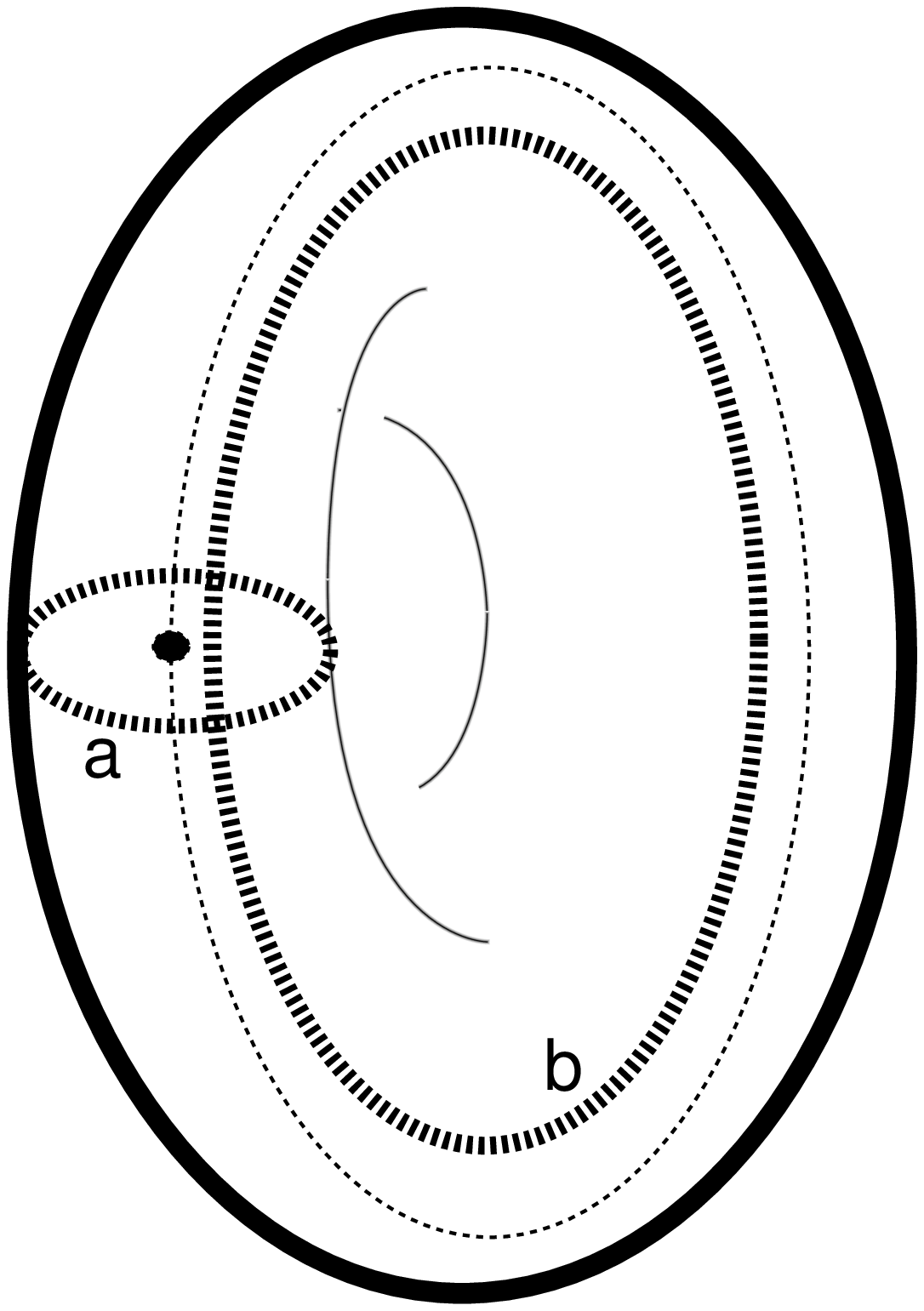}}

We now briefly consider invariants of links. We will consider links $\ell$
embedded in rational homology spheres $R$. Every knot in the link has a tubular
neighborhood which is a solid torus. One of the one-cycles in the boundary
of this solid torus is a parallel of the knot, while the other one-cycle is
the meridian of the knot. The complement of a link with $n$ components in a
rational homology sphere is a three-manifold with boundary $Y=R\setminus
\ell$, and
with $b_1(Y)=n$. The generators of the free abelian
group $H$ are the meridians of the link components. The Reidemeister-Milnor
torsion or Alexander polynomial of a link will be defined as the
Reidemeister-Milnor
torsion or Alexander polynomial of the manifold $Y$. In addition, one defines
the multivariable, symmetric Conway function
 of a link $\ell \subset R$ as
\foot{For $n>1$ $\nabla_\ell$ is a Laurent polynomial.}
\eqn\multial{
\nabla_{\ell} (t_1, \dots, t_n)= \tau (R \setminus \ell; t_1, \dots, t_n).}

Finally, we will need the one-variable Conway polynomial
$\nabla_{\ell}(z)$, where
$z=t^{1/2}- t^{-1/2}$. It is well known that this polynomial can be easily
computed from a simple set of skein relations and is related to the Milnor
torsion as follows (see \lescop, 2.3.13):
\eqn\rel{
|H_1 (R, \IZ)|  \nabla_{\ell}(z) = (t^{1/2}- t^{-1/2}) \tau (R \setminus
\ell; t, \dots, t),}
where $|H_1 (R, \IZ)|$ is the order of the finite group $H_1 (R, \IZ)$.
Notice, using \torso, that if $R= {\bf S}^3$, the Alexander polynomial of a
knot is equal to its Conway polynomial after the change of variable
$z=t^{1/2} - t^{-1/2}$. In general, the Conway
polynomial of an $n$-component link has the structure (see for instance
\lick, Proposition 8.7):
\eqn\alexconway{
\nabla_{\ell} (z)= z^{n-1} (a_0 + a_1 z^2 + \dots + a_d z^{2d}).}

\bigskip
\noindent
{\bf Remark.} In the above discussion we were
somewhat imprecise about the nature of the
Laurent series $\tau(Y;t_i)$. The precise story is
the following.
We can form two rings starting from $H$, $\IZ[H]$ and $\IZ[[H]]$. The
elements of $\IZ[H]$ are the Laurent polynomials $\IZ[t_1, t_1^{-1},
\dots, t_{b_1}, t_{b_1}^{-1}]$, where formally $t_i= \exp (h_i)$. $\IZ[H]$
is called the integral group ring of $H$, and can be understood as the set
of finitely supported $\IZ$-valued functions on $H$. We can introduce a
multiplicative action of $H$ on $\IZ[H]$ as follows: if $h \in H$ is
written as $h=\sum n_i h_i$, where $n_i \in \IZ$, and $f(t_1, \dots,
t_{b_1}) \in \IZ[H]$, then $h\cdot f(t_1, \dots, t_{b_1})=t_1^{n_1} \dots
t_{b_1}^{n_{b_1}} f(t_1, \dots, t_{b_1})$. The orbits of this
action are $\IZ[H]/H$, and the elements of this quotient group are elements
of $\IZ[H]$ defined up to multiplication by an element of the form
$t_1^{n_1} \dots t_{b_1}^{n_{b_1}}$. The reason to define this quotient
group is that the invariants we will consider, namely the Alexander
polynomial and the torsion of a three-manifold, are usually defined up to
multiplication by an element of this form (this is already the case, for
example, for the Alexander polynomial of a knot; see for instance \lick.)
Notice that there is a natural conjugation operation on $\IZ[H]$
given by $t_i \leftrightarrow t^{-1}_i$. The elements of $\IZ[[H]]$ can be
understood as formal series in the variables $t_i$, $t_i^{-1}$, therefore
$\IZ[[H]]$ is the set of $\IZ$-valued functions on $H$ (not necessarily with a
finite support). It can also be identified to the field of fractions of
$\IZ[H]$. The Reidemeister-Milnor torsion is
an element in $\IZ [[H]]/H$. It can be shown that it always
has a symmetric representative under $t_i \leftrightarrow
t_i^{-1}$, and throughout
this paper we will always work with symmetric representatives.
Similarly, the Alexander polynomials $\Delta_Y(t_i)$ is
properly thought of as an element of  $\IZ[H]/H$.
It too has a symmetric representative with which
we will always work.

\newsec{Seiberg-Witten invariants of $Y \times {\bf S}^1$}

In this section we will consider the SW invariants of four-manifolds with
the structure $X=Y \times {\bf S}^1$, and $b_1(Y)>0$. We first show using
simple arguments that the SW invariants of $X$ are determined by the
three-dimensional SW invariants of $Y$.
\foot{A more rigorous argument would follow the lines of \mst, where a
similar situation was analyzed involving the three-dimensional monopole
equations on $\Sigma_g \times {\bf S}^1$, $\Sigma_g$ being a Riemann surface.}
 Then we describe the relation of the three-dimensional SW invariants to
the Reidemeister-Milnor torsion, using the results of Meng and Taubes \mt.

\subsec{Seiberg-Witten invariants in three and
four dimensions}

The Seiberg-Witten equations and invariants can
be defined for both four-manifolds $X$ and
three-manifolds $Y$. The SW monopole
equations \monopole\ involve a pair $(A, M)$ consisting of
a connection and a section $M$
of a ${\rm Spin}^c$ line bundle
$c$. These equations can be written schematically as:
\eqn\threed{
\eqalign{
F(A)&={\overline M} \Gamma M  \cr
\Dsl_{A}M&=0.\cr} }

In addition to the four-dimensional version, the
three dimensional equations have been extensively studied over the past few
years both from the point of view of
mathematics \marwang\threekron, and also from the point of view of quantum
field theory \carey.
The available ${\rm Spin}^c$ structures are
determined as follows.
Every three-dimensional manifold is a ${\rm Spin}$ manifold \ms.
The ${\rm Spin}^c$ structures on $Y$ are simply $U(2)$ bundles $W_c$, with
first Chern class  $c_1(c)\in  H^2(Y, \IZ)$.
The ${\rm Spin}^c$ structure is uniquely determined
by its first Chern class $c_1(c)$.
\foot{After we have fixed an origin for the action of  the
$2$-torsion elements  on the ${\rm Spin}^c$ structures.
Technically, the set of ${\rm Spin}^c$ structures is
a ``torsor.''}
Since $Y$ is Spin, $c_1(c)$ is always divisible by $2$ in $H^2(Y, \IZ)$.
Here and
below we will  denote the projection of $c_1(c)$ into  the free abelian
group $H(Y)=H^2 (Y, \IZ)/{\rm Tor}(H^2 (Y, \IZ))$ by $\bar c_1 (c)=x$.
As in four dimensions, one now defines the three-dimensional SW invariant,
denoted by $SW (Y, c)$,  as a signed sum of solutions modulo gauge equivalence
to the  equations \threed.

Let us now consider  the relation between
the SW invariants on a three-manifold $Y$ and those on the
associated four-manifold $X= Y \times {\bf S}^1$,
 where $Y$ is a compact, oriented three-manifold with $b_1(Y)>0$. The
numerical invariants
are related by $\chi(X)= \sigma (X) =0$, $b_1(X)=b_1(Y)+1$.
We also have $b_2^+ (X)= b_1 (Y)$.

As for the SW invariants
we will now show that the ${\rm Spin}^c$ structures
of $X$ and $Y$ supporting nonzero invariants are
essentially the same.
Since $Y$ is spin, so is $X$.
 The first thing to notice is that, even when $b_1 (Y)=1$, there is no
wall-crossing (WC) for the SW invariants of $X$, provided
that they are computed with a small perturbation of the equations. (There
is wall-crossing for the
stable SW invariants computed with a very large perturbation, and this will
be considered presently.)
The reason for this is that non-trivial SW invariants arise only when the
dimension of the moduli
space associated to a ${\rm Spin}^c$-structure with determinant line bundle
$2\lambda$ is $d_{\lambda}=\lambda^2 \ge 0$ (recall that $\sigma (X) = \chi
(X)=0$), but for small perturbations of the SW equations a necessary
condition for WC is $\lambda^2<0$. Therefore, we can choose any
metric on $Y \times {\bf S}^1$ to compute the invariants. We will work with
a product metric corresponding to a radius $R$ for the circle, {\it i.e.}
$g=R^2 d \varphi^2 + h_{ij}dx^i dx^j$, where
$\varphi$, $x^i$ are coordinates for the circle and $Y$, respectively, and
$h_{ij}$ is an arbitrary
Riemannian metric on $Y$.
The four-dimensional Dirac equation equation $\Gamma^{\mu}D_{\mu}M=0$,
$\mu= \varphi, i$ can be written in local coordinates as
\eqn\dirac{
{1 \over R} \Gamma^{\varphi} D_{\varphi} M +\Gamma^i D_i M=0,}
where $(\Gamma^{\varphi})^2=1$. If we now consider the limit $R\rightarrow 0$,
we see that the monopole field has to be covariantly constant in the
direction of
the circle. Similarly, the other monopole equation $F^+ - \overline M
\Gamma M=0$
implies in this limit that $F_{i \varphi}=0$. This means that the Chern class
of the ${\rm Spin}^c$ structure is a two-cohomology class on $Y$, in other
words, that the monopole equations will
have a solution only if the ${\rm Spin}^c$-structure on $X$ is the pullback
of a
${\rm Spin}^c$-structure on $Y$. Therefore, the pair $(A,M)$ is induced (up
to a gauge transformation) by a pair $(A,M)$ on $Y$, and what we have to solve
are precisely the monopole equations that arise on the three-dimensional
manifold $Y$ by dimensional reduction.

One can in fact prove that the SW invariants for the manifold $X$ are given
by \threekron
\eqn\useful{
SW(Y\times {\bf S}^1, \pi^{*} (c))= SW(Y,c).}
where the ${\rm Spin}^c$-structure on $X$ is the pullback $\pi^* (c)$
of a ${\rm Spin}^c$-structure $c$ on $Y$. As we have seen, the other ${\rm
Spin}^c$ structures on $X$ have a zero invariant. This means that the
four-manifold $X=Y \times {\bf S}^1$ is of simple type, as
the only possible basic classes satisfy $x^2=0$.

\subsec{The Meng-Taubes result}
As we have seen, the problem of computing the SW invariants of the
four-manifold
$Y \times {\bf S}^1$ reduces to the problem of computing the 3D SW
invariants of the manifold $Y$. A very interesting structural
result for these invariants has been found by Meng and Taubes \mt.

\subsubsec{$b_1 (Y)>1$}

First we describe the result
for three-manifolds with $b_1(Y)>1$. We define the Seiberg-Witten polynomial
${\underline {SW}} (t_i)\in \IZ[H]/H$ as follows.
Let $x= \sum_{i=1}^{b_1} 2k_i h_i$ be the free part of the first Chern
class of a ${\rm Spin}^c$-structure $c$ on $Y$, where $k_i$ are integers
and $h_i$ are generators of $H=H^2 (Y, \IZ)/{\rm Tor}(H^2(Y, \IZ))$, and
let $SW(Y,c)$ be the
corresponding SW invariant. The SW series of $Y$ is defined as
\eqn\mtseries{
{\underline {SW}} (t_i) = \sum_{x \in H} \Biggl(\sum_{c|\bar
c_1(c)=x}SW(Y,c) \Biggr) t_1^{k_1} \dots t_{b_1}^{k_{b_1}},}
where, for a given $x$ in $H$, we sum over all the SW invariants of the
$|{\rm Tor}H^2(Y, \IZ)|$ ${\rm Spin}^c$-structures $c$ with the same $x$.
A different choice of generators gives a different representative in the
orbit. In the orbit of ${\underline {SW}}$ there is always an element which
is invariant
under conjugation. This is a consequence of the charge-conjugation
invariance of the SW equations \monopole.
The main result of Meng-Taubes is that
\eqn\swmilnor{
{\underline {SW}} (t_i)= \tau (Y;t_i),
}
where $\tau (Y;t_i)$ is the Milnor torsion. In this case, both elements are in
$\IZ[H]$, {\it i.e.} they are polynomials in the $t_i$, $t_i^{-1}$
variables. On the SW side, this is a simple consequence of the fact that
there are only a
finite number of basic classes.

As a simple example of \swmilnor, one can consider the three-manifold
$Y=\Sigma_g \times {\bf S}^1$, where $\Sigma_g$ is a Riemann surface of
genus $g$. In this case, the torsion is $\tau (\Sigma_g \times {\bf
S}^1;t_i)=(t_1^{1/2}- t_1^{-1/2})^{2g-2}$, where $t_1={\rm e}^{2[{\bf
S}^1]}$. This is precisely the SW series of the manifold $\Sigma_g \times
{\bf T}^2$, and using \useful\ we see that \swmilnor\ holds.

\subsubsec{$b_1 (Y)=1$}
We now consider the structure of the invariants when $b_1 (Y)=1$. In this
case, the SW invariants have a mild dependence on the perturbation of the
equations, and this
amounts to a choice of a generator of $H^1(Y, \IZ)$ \mt. Notice that, as
$H^1(Y, \IZ)\simeq \IZ$, there are only two possible generators that differ
in their sign.
\foot{Note in particular that $H^1(Y, \IZ)$ has no
torsion, in contrast to $H^2(Y,\IZ)$.}
To understand this dependence on the perturbation, it is useful to consider
a four-dimensional version of this story, focusing again on the four manifold
$X= Y \times {\bf S}^1$, with $b_1(Y)=1$. The structure of the cohomology
ring of $X$ is the following:  We denote by $S_2$ a generator of $H_2(Y,
\IZ)$ and by $\gamma$ a generator of
$H_1(Y,\IZ)/{\rm Tor}(H_1(Y, \IZ))$. Then there is another two-cycle in $X$
given by cup product
$S_1=\gamma \cup {\bf S}^1$, where ${\bf S}^1$ denotes here a generator of
$H_1({\bf S}^1)$.
We also have   cohomology classes $\alpha_Y \in H^2(Y, \IZ)/{\rm Tor}
(H^2(Y, \IZ))$
and $\beta_Y \in H^1(Y, \IZ)\simeq \IZ$, and such that
\eqn\pduals{
[S_1]= \alpha_Y, \,\,\,\,\,\ [S_2] = \beta_Y \wedge e,}
where $e=d\varphi  $ is a generator of $H^1({\bf S}^1, \IZ)$ and the
brackets mean Poincar\'e duals. The intersection form of
$X$ is the even unimodular lattice $ II^{1,1}$,
and $[S_1] \cdot [S_2] =1$.

If we consider the usual SW invariants (with a very small perturbation),
there will be no wall-crossing, as we remarked above. Assume now that we
turn on a perturbation, and we consider the ``twisted" SW equations on $X$
\monopole\okonek
\eqn\twisted{
(F+2 \pi i \beta)_+={\overline M}M,}
where $\beta$ is a closed two-form which represents $b\in H^2(X, \IR)$.
Equivalently, we consider a ${\rm Spin}^c$-structure such that the free
part of its first Chern class is $x-b$.

\ifig\taubesch{Chamber structure in the space of perturbations
and line bundles for the case $b_1(Y)=1$. The perturbation
has cohomology class $b=2r \alpha_Y$ and the line bundle
has first Chern class $x=2k \alpha_Y$. When $k>0$ the
number of solutions of \twisted\ jumps by $+k$ as
we cross from chamber $\CC^-$ to $\CC^+$.  }
{\epsfxsize2.5in\epsfbox{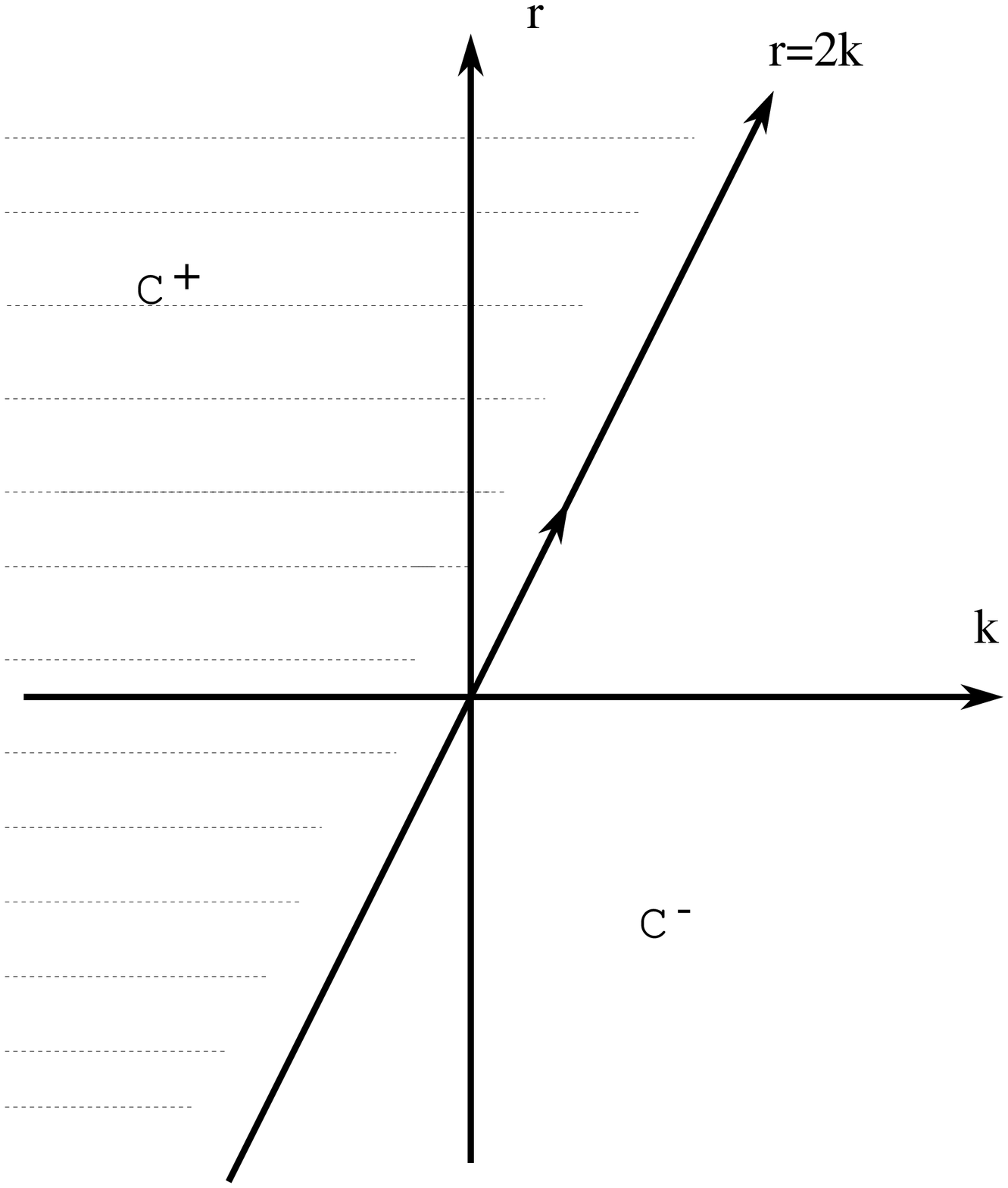}}

There is now a chamber structure in the space of perturbations, where the
walls are defined by
\eqn\wcgeneral{
x=b. }
We will choose the perturbation in such a way that $b=2r \alpha_Y$, where
$r$ is a real number.
Following the arguments in the previous subsection, we again see that the
four-dimensional SW invariants will be zero for ${\rm Spin}^c$-structures
which are not the pullback of a ${\rm Spin}^c$ structure on $Y$. If the
${\rm Spin}^c$-structure is the pullback of  a ${\rm Spin}^c$-structure $c$
on $Y$, with $x$ of the form $x=2k \alpha_Y$, the corresponding SW
invariants will be the SW invariants computed from the perturbed 3d
monopole equation
\eqn\tdper{
F(A)={\overline M} \Gamma M-2\pi i r \alpha_Y,}
for the ${\rm Spin}^c$-structure $c$ on $Y$.
As the walls are given by \wcgeneral, the chambers are \eqn\cpchambers{
{\cal C}^{+}= \{ b=r\alpha_Y \in H^2(Y,\IR): r>2k \}, \,\,\,\,\,\  {\cal
C}^{-}= \{ b=r\alpha_Y \in H^2(Y,\IR): r<2k \}.}
and are illustrated in \taubesch.

The perturbation in \twisted\ has a very natural interpretation in terms of
the underlying physical theory (at least in the K\"ahler case): when the
twisted theory is perturbed by a mass term that breaks ${\cal N}=2$
supersymmetry down to ${\cal N}=1$, as in \wittk, the mass parameter must
be a self-dual two-form . In the effective theory,   this perturbation
modifies the usual Seiberg-Witten monopole equations and gives \twisted.

We can now define the ``stable" SW invariants $SW^{\pm}(Y,c)$ in such a way
that, for every ${\rm Spin}^c$-structure $c$, the SW invariant is computed
in the chamber ${\cal C}^{\pm}$,
{\it i.e.} after turning on a perturbation with $r$ sufficiently large. It
is easy to see that the choice of a chamber is equivalent to a choice of a
generator for $H^1(Y,\IZ)$ \mt. Let $o=\pm \beta_Y$ be a generator of
$H^1(Y,\IZ)$. If we write the condition on the
perturbation as \mt
\eqn\choicepert{
\int_Y b\wedge o \quad > \int_Y c_1(c) \wedge o,}
we see that  the choices $o=\pm \beta_Y$ give the two chambers ${\cal
C}^{\pm}$, {\it i.e.},
$r>2k$ and $r<2k$.  Using these stable invariants
we can define a series just as in \mtseries:
\eqn\cpseries{
{\underline {SW}}^{\pm} = \sum_{x \in H} \Biggl(\sum_{c|\bar
c_1(c)=x}SW^{\pm}(Y,c) \Biggr) t^{k} .} Notice that
in the stable case there are infinitely many characteristic elements with a
nonzero
SW invariant. This is a consequence of the definition of stability, which
involves an arbitrarily large perturbation of the equations. In fact, the
effect of considering the stable invariants is to add an (infinite)
universal series to the polynomial defined by the usual Seiberg-Witten
invariants $SW(Y,c)$ (computed
for a small perturbation). This can be seen as follows: let $c$ be a ${\rm
Spin}^c$-structure with $x=\bar c_1(c)=2k \alpha_Y$, and let's consider the
stable SW invariants associated to the chamber ${\cal C}^+$. If $k<0$, a
small perturbation is already in the chamber ${\cal C}^+$, so
$SW^+(Y,c)=SW(Y,c)$.
When $k>0$, the difference between the invariants is the wall-crossing
term. The wall-crossing formula for a nonsimply connected four-manifold $X$
has been obtained in \okonek\liliu, and rederived in \ns\ using the
$u$-plane integral. In general, one has:
\eqn\wcthree{
SW^+ (c)-SW^-(c)= {1 \over 2} (-1)^{b_1} (c_1 (c),\Sigma) {\rm vol }({\bf
T}^{b_1}),
}
where $\Sigma$ is a certain two-dimensional cohomology class in $X$, and
${\bf T}^{b_1}$ is the torus $H_1(X, \IR)/H$ (see \ns\ for the details).
Notice that all the ${\rm Spin}^c$-structures with the same $x=\bar c_1
(c)$ have the same wall-crossing behavior. In our case, the four-manifold
$X=Y \times {\bf S}^1$ has $b_1(X)=2$, and the two-dimensional class
is given by
\eqn\susigma{
\Sigma=\beta_Y \wedge e = [S_2] , }
while $x=2k [S_1]$. Therefore,
using \wcthree, we obtain, for $k>0$,
\eqn\cross{
SW^+(Y,c)=SW(Y,c) + k.}
The universal term in ${\underline {SW}}$ is then given by
\eqn\univ{
|{\rm Tor}(H_1 (Y, \IZ))| \sum_{k=1}^{\infty} kt^k =  {t \over (1-t)^2}
\cdot| {\rm Tor}(H_1 (Y, \IZ))| ,}
where we have included the factor $|{\rm Tor}(H^2(Y, \IZ))|=|{\rm Tor}(H_1
(Y, \IZ))|$ accounting for the different ${\rm Spin}^c$-structures with the
same $\bar c_1 (c)$.
A similar computation shows that one obtains the same universal
contribution for the Seiberg-Witten invariants computed in the other
chamber ({\it i.e.} although the individual
SW invariants differ, the formal series is the same). We then find,
\eqn\finwc{
{\underline {SW}}^{\pm} = {t \over (1-t)^2} \cdot| {\rm Tor}(H_1 (Y, \IZ))|
+ \sum_{x \in H} \Biggl(\sum_{c|\bar c_1(c)=x}SW(Y,c) \Biggr) t^{k},}
where the second term in the right hand side is a polynomial, since the SW
invariants appearing there are computed for a small perturbation.

The result of Meng and Taubes when $b_1(Y)=1$ is that
\eqn\mtone{
{\underline {SW}}^{\pm}=\tau(Y;t).}
Comparing \milnor\ with \finwc\ we see that the first term in the
Reidemeister-Milnor torsion is the  sum of the infinite series of
wall-crossing terms. According to \mtone\ the remaining piece of the
torsion in \milnor, which is a polynomial, must be identified with the SW
invariants computed for a small
perturbation. Equating the coefficients of the two polynomials in \mtone\
and \finwc, one finds:
\eqn\zeroper{
\sum_{c| \bar c_1(c)=2k \alpha_Y} SW(Y,c)= \sum_{j=1}^{r-|k|} j a_{j+|k|}, }
in accordance with \threekron.

\newsec{Donaldson-Witten partition function for $b_1(Y)>1$}

In the next two sections we combine the
facts reviewed above
with the formulae for the Donaldson-Witten
partition function on four-manifolds
of the type $X=Y \times {\bf S}^1$.
The extension of the Donaldson-Witten function to
non-simply connected four-manifolds was begun in
\mw\lns. It was completed in \ns. Somewhat later
the same result was stated, less precisely,
in \munnsc.

The Donaldson-Witten  partition function
on a four-manifold $X$ depends crucially
on $b_2^+(X)$. As we have noted,  the  four-manifolds $X=Y \times {\bf
S}^1$  have
\eqn\bplus{
b_2^+(X)=b_1(Y)
}
and hence $Z_{DW}(X)$ depends crucially on $b_1(Y)$.
For $b_1(Y)>1$, the only contribution to the Donaldson-Witten function
comes from the SW contributions. It was explicitly verified in \ns\ that
Witten's formula for the Donaldson invariants of simple type manifolds
\monopole\ is modified in the nonsimply connected case only when one and
three-observables are included. The result for the partition function is then,
\eqn\wittens{
Z_{DW}(Y \times {\bf S}^1)=2 (1+ i^{-w_2^2(E)})
\sum_{\lambda} e^{2i\pi(\lambda_0\cdot\lambda+\lambda_0^2)}SW(\lambda). }
Here the sum on $\lambda$ is a sum over
${\rm Spin}^c$ structures with $c_1(c)=2\lambda$,
  $w_2(E)$ is an integer lift of the Stiefel-Whitney class of the bundle,
and $2 \lambda_0=w_2(E) $. For  $w_2(E)=0$, one has to evaluate the sum
over all the ${\rm Spin}^c$ structures, and using \mtseries\swmilnor\ and
\torgo\ this is just $\Delta_Y(1, \dots, 1)$. We then obtain,
\eqn\simple{
Z_{DW}(Y\times {\bf S}^1) = 4 \Delta_Y(1, \dots, 1),}
which is simply the sum of all the coefficients of the Reidemeister-Milnor
torsion.

\subsec{Relation to the Casson-Walker-Lescop invariant}

We will now relate this result to   Lescop's extension of the
Casson invariant \lescop, which we will
refer to as the Casson-Walker-Lescop invariant
$\lambda_{CWL}(Y)$.
 First, we will recall the surgery operation on a link $\ell$
 embedded in a rational homology sphere $R$ to obtain a new three-manifold.
 To perform surgery on a link, we need to associate to any knot $K_i$
 in the link $\ell$ a satellite $\mu_i$, {\it i.e.} a closed curve on
 the boundary $\partial (T(K_i))$ of a tubular neighborhood $T(K_i)$
 of $K_i$. The homology class of the satellite in the boundary is
 specified by two rational numbers $(p_i, q_i)$ that determine the
 framing of the knot. $p_i$ is the linking number ${\rm lk}
 (\mu_i, K_i)$ ({\it i.e.} the number of ``twists" between
 the knot and its satellite). If we consider the homology class
 of the satellite in the solid torus, we have that $[\mu_i]=q_i [K_i]$.
\foot{The  $q_i$ are always integers.  If $R=S^3$ then the
$p_i$ are also integers. In general the $p_i$ can be
rational.}
When $R={\bf S}^3$, the $(p_i, q_i)$ are simply the number of
times that the satellite wraps the meridian and parallel,
respectively, of the knot $K_i$. The quotient $p_i/q_i$ is the surgery
coefficient of the component $K_i$. To do the surgery on $\ell$,
we consider solid tori $(D^2 \times {\bf S}^1)_i$ together with
the homeomorphism $h_i$ that sends the meridian $({\bf S}^1)_i$
to the satellite $\mu_i$ on $\partial(T(K_i))$. We then cut the
solid tori $T(K_i)$ around each of the knots and glue the
$(D^2 \times {\bf S}^1)_i$ through the homeomorphisms $h_i$.
In this way we obtain a new three manifold obtained from $R$
through the above surgery presentation. The linking matrix of
the presentation is the matrix $L_{ij}$ with components ${\rm lk}
(K_i, K_j)$ when $i \not= j$, and $p_i/q_i$ when $i=j$.

In order to relate \simple\ to Lescop's extension of the Casson
invariant, the key fact is that any oriented, closed three-manifold
$Y$ can be obtained by surgery on a link in a rational homology
sphere (see \lescop, Lemma 5.1.1). This link has $b_1(Y)$
components and its linking matrix can be chosen to be
null ({\it i.e.}, all its entries are zero). We also
have that $|H_1(R, \IZ)|=|{\rm Tor} (H_1 (Y, \IZ))|$.

\ifig\borromean{One presentation of the Borromean
link. Note that no two components are linked.  }
{\epsfxsize2.5in\epsfbox{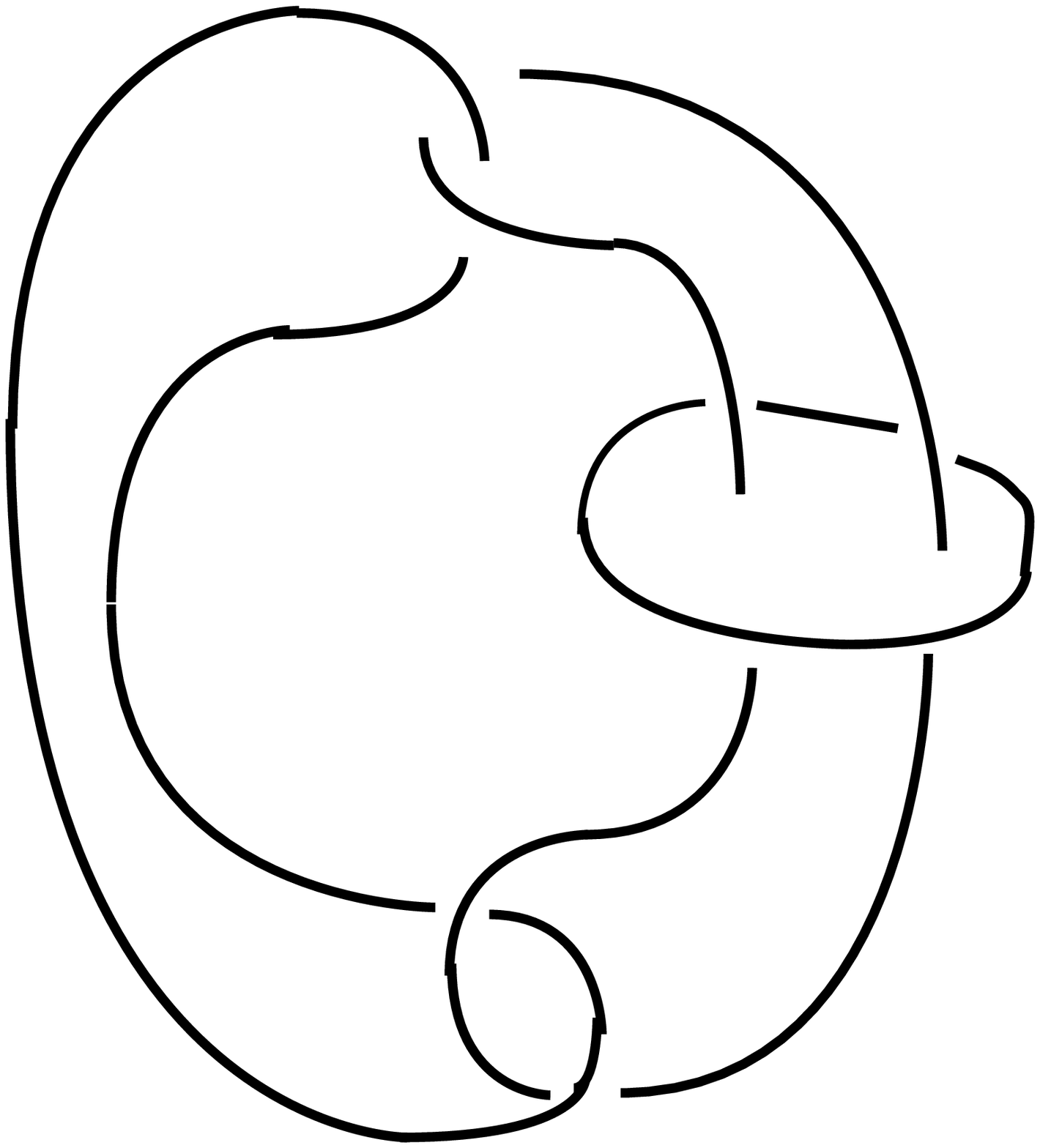}}

As an example of this, consider the Borromean link
in ${\bf S}^3$, which has thee components. The linking numbers
of its components are zero. Performing surgery on this link,
with $p_i/q_i=0$, we obtain the three torus ${\bf T}^3$!

The links considered in these surgery
presentations, have the property  that the linking number of any two
components is zero. Such links are called homology unlinks. For a
homology unlink  the coefficient $a_0$ in \alexconway\ vanishes
(see \lescop, Remark 5.2.8). Therefore, for
homology unlinks the Conway polynomial has
the structure:
\eqn\part{
\nabla_{\ell}(z)=z^{b_1+1} (a_1 + a_2 z^2+ \dots + a_d z^{2(d-1)}),}
where $z=t^{1/2}-t^{-1/2}$ and $b_1$ is the number of components of the link.
On the other hand, the torsion of $Y$ can be easily computed in terms of the
multivariable Conway function of the link, when the linking
matrix of the surgery presentation is null. The relation
between them is (see \turaev, section 4.3.4, Remark 2):
\eqn\spal{
\tau (Y; t_1, \dots, t_{b_1})= \biggl( \prod_{i=1}^{b_1}
(t_i^{1/2} - t_i^{-1/2}) \biggr)^{-1} \nabla_{\ell} (t_1, \dots, t_{b_1} ).
}
As a consequence of \spal, notice that, if the manifold
$Y$ is obtained by $0$-surgery on a knot
$K \subset {\bf S}^3$, the Alexander polynomial
of the manifold $Y$ is the Alexander polynomial of the knot $K$.
As an example of \spal\ for links, consider the Borromean link in \borromean.
This link has the multivariable Conway polynomial
$\nabla_l (t_1, t_2, t_3) = \prod_{i=1}^3(t_i^{1/2}-t_i^{-1/2})$,
therefore the torsion of ${\bf T}^3$ is $\tau({\bf T}^3)=1$.

Returning to the general case, and using
  now \rel, \multial, \part\ and \spal, we obtain for $b_1 (Y)>1$
\eqn\relation{
\Delta_Y (1, \dots, 1) = |H_1 (R, \IZ)| a_1.}
The right hand side of this expression is precisely
Lescop's extension of the Casson invariant
for manifolds with $b_1(Y)>1$ (see \lescop, 5.1.7), therefore we have
\eqn\finalone{
Z_{DW}(Y\times {\bf S}^1) = 4\lambda_{CWL} (Y).}
The factor of four arises as follows. The monopole and
dyon cusps contribute equally. The other factor of
2 comes from the center of the gauge group $SU(2)$.

\subsec{Relation to Floer homology}

As we mentioned in the introduction, one of our motivations to analyze the
partition function of
Donaldson-Witten theory for manifolds with the structure $Y \times {\bf S}^1$
was to obtain a relation with the Euler characteristic of the Floer
cohomology of $Y$. One can check these expectations by considering
the three-manifolds $Y_g= \Sigma_g \times {\bf S}^1$, where $\Sigma_g$ is a
Riemann surface of genus $g$. The first Betti number for this class of
manifolds
is $b_1(Y_g)= 1+2g$. The Floer cohomology of these manifolds is
computed by turning on a non zero flux on $\Sigma_g$, {\it i.e.}
$w_2 (E)=[{\bf S}^1]$. In this case, the expressions \simple\finalone\
remain valid, as \mst\ the basic classes on $Y_g$ are two-dimensional
classes on $\Sigma_g$ and they have no intersection with $w_2(E)$.
The ring structure of the Floer cohomology for these manifolds is
known \munozfloer\ and in particular $\chi (HF (Y_g))=0$ except for
$g=1$, where one has $\chi (HF ({\bf S}^1)^3))=1$. This is in perfect
agreement with the behavior of the Casson invariant, which vanishes for
$b_1(Y)>3$, and
has $\lambda_{CWL} (({\bf S}^1)^3)=1$ \lescop. We then see that the
partition function should be related to the Euler characteristic of
the Floer cohomology, for manifolds with $b_1(Y)>1$, as
\eqn\relfloer{
 Z_{DW}(Y\times {\bf S}^1) = 4 \chi (HF(Y)).}

\subsec{Extension to higher rank}
When $b_1 (Y)>1$, the partition function of Donaldson-Witten theory
for gauge group $SU(N)$ can be easily computed using the results of \mmtwo.
In this paper, a simple expression for the $SU(N)$ Donaldson-Witten function
on manifolds with $b_2^+>1$ and of simple type was derived using the $u$-plane
approach of \mw.
This expression is given in equation (9.17) of \mmtwo. For the partition
function,
one obtains the following equation, which generalizes \wittens:
\eqn\simple{
\eqalign{
Z_{DW}^{SU(N)}(X) &= N  {\widetilde \alpha}_N^\chi
{\widetilde \beta}_N^\sigma \sum_{k=0}^{N-1} \omega^{k[(N^2-1)
\delta + N \vec \lambda_0\cdot \vec \lambda_0]} \cr
 & \,\,\,\,\,\,\,\, \cdot \sum_{\lambda^I} {\rm e}^{ 2\pi
i (\lambda^I, \lambda_0^I) }  \Bigl(\prod_{I=1}^{N-1}
SW(\lambda^I)\Bigr)\prod_{1\le I< J \le r}
 q_{IJ}^{-(\lambda^I, \lambda^J)}.\cr}
}
In this equation, $\widetilde \alpha_N$, $\widetilde \beta_N$ are
constants, $\omega =
\exp [ i \pi /N]$, $\vec \lambda_0$ is an integral lifting of the generalized
Stiefel-Whitney class of the gauge bundle (see \mmtwo\ for details),
$r=N-1$ is the rank of the gauge group, and $\delta =(\chi + \sigma)/4$.
The terms
$q_{IJ}$ are the leading terms of the off-diagonal couplings. We have
also included an overall $N$ factor corresponding to the order of the center
of the gauge group. Finally, the sum over
$k$ is a sum over the ${\cal N}=1$ vacua.

If we consider a manifold $X=Y \times {\bf S}^1$, with $b_1(Y)>1$, and we
choose
$\vec \lambda_0=0$, the above
expression factorizes completely, as the exponents of the nondiagonal couplings
$q_{IJ}$ are zero. We then find
\eqn\cassun{
Z_{DW}^{SU(N)}(Y \times {\bf S}^1) = N^2 \bigl( \lambda_{CWL}(Y) \bigr)
^{N-1},}
which generalizes \finalone\ to $SU(N)$. It would
be very interesting to compare \cassun\ with the generalizations of the Casson
invariant that can be obtained using Rozansky-Witten invariants. \foot{
Investigating generalizations of the Casson invariant for other gauge groups
using
Rozansky-Witten theory has also been recently proposed by
G. Thompson in \gthompson.}
These generalizations might be more nontrivial for
the case $b_1(Y)\leq 1$.

\newsec{Three-manifolds with $b_1(Y)=1$}

\subsec{The Donaldson-Witten partition function}

When the three-manifold $Y$ has $b_1(Y)=1$, the four-manifold $X=Y \times
{\bf S}^1$ has
$b_2^+ (X)=1$. This means that we have to take into account both the SW
and the
$u$-plane contribution, as explained in \mw. The $u$-plane contribution for
nonsimply connected manifolds has been analyzed in detail in \ns. We also
have to take into account that
the Donaldson-Witten function depends now on the period point of the metric,
and in general
our answers will be metric-dependent. We will consider in particular
the chambers corresponding to a small or big radius for the circle ${\bf S}^1$.

When studying the relation between Donaldson invariants and
three-dimensional invariants, it
is important to take into account the torsion in $H^2 (X, \IZ)$. The
inclusion of torsion in the $u$-plane integral can be done in the
following way: the partition function
of the photon includes a sum over topological sectors, {\it i.e.}
over topological classes of line
bundles. This means that we have to sum over torsion classes as
well in $H^2 (X, \IZ)$. But the
photon partition function
depends on the topology of the
gauge (line) bundle only through the
curvature $2$-form $F_A$   and therefore
  is only sensitive to the torsion-free part of $H^2 (X, \IZ)$. This means
that, when summing over
all the topological sectors, we
will have a sum over the classes in $H^2 (X, \IZ)/
{\rm Tor}(H^2(X, \IZ))$ and then include
a global factor $|{\rm Tor}(H^2(X, \IZ))|$ multiplying the $u$-plane integral.
\foot{One could define other topological theories
by introducing a nontrivial character of the
group ${\rm Tor}(H^2(X, \IZ))$ into the path integral.
This would be an analog for Donaldson-Witten theory
of ``discrete torsion'' of a  string theory orbifold. We will
not investigate this possibility further here.}
In particular, the wall-crossing formula will have this factor in the
presence of torsion,
as noticed in \kots. When matching to SW wall-crossing as in \mw\ns, this
factor will be
present as well on the SW side, as the WC crossing of the SW invariants
only depends on the
torsion-free part of the ${\rm Spin}^c$ structure \okonek\liliu.

We can now compute the $u$-plane contribution to the Donaldson-Witten
function of manifolds
$X=Y \times {\bf S}^1$. Let's first analyze the metric dependence. A
generic period point has the structure,
\eqn\period{
\omega={1 \over {\sqrt 2}} ({\rm e}^{\theta} [S_1] + {\rm e}^{-\theta}
[S_2] ).}
The limit of a small radius for ${\bf S}^1$ corresponds to $\theta
\rightarrow \infty$, as the volume
of the $S_1$ is
\eqn\volume{
\int_{S_1} \omega = {1 \over {\sqrt 2}} {\rm e}^{-\theta}.}
The other limit, $\theta \rightarrow - \infty$, corresponds to a large
radius for the circle.
It is helpful to keep the following example in mind.
Suppose $Y$ is a circle bundle over $\IC P^1$, and
$\alpha_Y$ is the volume form of a metric
$ds^2$ on $\IC P^1$ normalized
to unit volume. Then we could consider a  metric on
$X$ given by:
\eqn\helpful{
  R^2 (d \varphi)^2 + (d \psi)^2 + { 1 \over  R^2} ds^2
}
where $\psi$ is a coordinate on the fiber. Thus
  we can identify $R$ the radius of the circle
parametrized by $0 \leq \varphi \leq 1$ with \volume.

We first analyze the Donaldson-Witten partition function in the limit
$R\rightarrow 0$, and
with no magnetic fluxes, so we put $w_2(E)=0$. In such a situation, the
$u$-plane integral can be computed
directly, as in section 8 of \mw. For  $R\rightarrow 0$, the right
choice of the reduction vector $z$ is in this case
\eqn\zetared{
z=[S_1], \,\,\,\,\,\ z_+^2={1 \over 2} {\rm e}^{-2\theta} \ll 1.}
Again, if $Y={\bf S}^2 \times {\bf S}^1$, $X=\IC P^1 \times {\bf T}^2$,
this is the chamber where the volume of the torus is very small. As
our manifold is non-simply connected, we have to use the expressions
of \ns. These involve a
choice of cohomology class $\Sigma$ and a
modified two-observable $I(\tilde S)$.  In our case
the cohomology class $\Sigma$ is given in \susigma\ and
the modified two-observable is obtained from
\eqn\stilde{
(\tilde S, z) = (S,z) - { \sqrt 2 \over 16} { d\tau \over du} \Omega,}
where $\Omega$ is the volume element of  the torus
${ \bf T}^2= H^1(X, \IR)/H^1(X, \IZ)$ \ns. The holomorphic function $f$
introduced in
\mw\ns\ is
\eqn\holof{
f= { {\sqrt 2} \over 8 \pi i } {du \over d \tau} \exp \biggl[ {\sqrt 2
\over 32} a {d\tau \over da}
(S, \Sigma) \Omega + S^2 T(u) \biggr] , }
where $du/d\tau$, $a$ and $T(u)$ are certain modular
forms described in \mw\ns.The $u$-plane integral in this chamber is given by:
\eqn\uplane{
Z_u = -4 {\sqrt 2} \pi i \cdot 2^{9 b_1/4} i^{b_1/2}
|{\rm Tor}(H_1(Y, \IZ))| \biggl\{ \sum_I  \biggl[  { f_I h_I  \over
1-{\rm e} ^{-i  (\tilde S, z)/ h_I }  } \biggr] _{q^0}  \biggr\} ,}
where $f_I, h_I$ are some more modular forms defined
in \mw\ns.
The sum is over   four regions at infinity of the
$u$-plane, labelled $I=(\infty, 0)$, $(\infty, 1)$, $(\infty, 2)$,
$(\infty, 3)$,
and the monopole and dyon regions of
the $u$-plane, labelled $I=M,D$. These regions are each a copy
of a fundamental domain for ${\rm SL}(2, \IZ)$. Together the
six regions form a fundamental domain for $\Gamma^0 (4)$.
This domain has three cusps: the cusp at infinity
(corresponding to the four regions $I=(\infty,0), \dots, (\infty,3)$)
and the regions near $\tau=0$ and $\tau=2$ (corresponding to $I=M,D$,
respectively.) The numerical prefactor involving $b_1$ in \uplane\
comes from the measure for the one-forms and was determined in \ns\
by comparing to known topological results.
Using the K\"unneth theorem and the universal coefficient
theorem we have
${\rm Tor} (H^2 (X, \IZ)) \cong {\rm Tor}(H_1(Y, \IZ))$
so  the prefactor for the torsion classes can be
written as
 $|{\rm Tor} (H^2 (X, \IZ))|= |{\rm Tor}(H_1(Y, \IZ))|$. As
 $f$ in \holof\ involves the volume element $\Omega$,
 we have to expand the functions appearing in \uplane\ in
$\Omega$ and then integrate. The computation is easy, and
one finds that each region $I$ contributes
$-2 |{\rm Tor}(H_1(Y, \IZ))|/12$. In conclusion, we   find
\eqn\total{
Z_u (Y \times {\bf S}^1) =-|{\rm Tor} H_1 (Y, \IZ)|.}

Let's now consider the SW contribution. The SW invariants,
computed for a small perturbation, do not depend on the metric.
To obtain the Donaldson-Witten partition function, we have
to sum over all SW invariants, as in \simple.
Using \alexander\ and \zeroper, we find
\eqn\sumspin{
\sum_{c} SW(Y,c) = \sum_{\ell=1}^r \ell^2 a_\ell = {1 \over 2} \Delta_Y''(1),}
where $\Delta_Y$ is the Alexander polynomial. Taking into account \torsum,
we can write the
partition function of Donaldson-Witten theory,
$Z_{DW}=Z_u + Z_{SW}$, in terms of the Alexander polynomial of $Y$ as
follows:
\eqn\casiles{
Z_{DW}(Y \times {\bf S}^1)= 2 \Delta_Y''(1) -\Delta_Y (1).}
It is interesting to compare this result with the
Casson invariant as extended by Lescop \lescop.
For manifolds with
$b_1(Y)=1$ it is given by (see \lescop, 5.1):
\eqn\boneles{
\lambda_{CWL} (Y)= {1 \over 2} \Delta_{Y}''(1) - {|{\rm Tor}(H_1(Y, \IZ))|
\over 12}.}
We therefore arrive at one of the key
results of this paper:
\eqn\comp{
Z_{DW} (Y \times {\bf S}^1)=4 \lambda_{CWL} (Y) - {4 \over 6}|{\rm
Tor}(H_1(Y, \IZ))| .}
Note that, even after accounting for a factor of $4$, as
in \finalone, the invariants do not agree. It is important to
notice that the result \casiles\ is obviously an integer,
while Lescop's extension of the Casson invariant for manifolds
with $b_1(Y)=1$ takes values in $\IZ /12$. For instance,
for $Y={\bf S}^2 \times {\bf S}^1$ (which has $\Delta_Y(t)=1$),
one has $\lambda_{CWL} (Y)=-1/12$, but $Z_{DW}
(Y \times {\bf S}^1) = -1$. We will comment on this
disagreement below, as well as on the relation of
\casiles\ to the results of \rw.

The fact that our result is an integer suggests that it is
related to the Euler characteristic of the Floer homology
of $Y$. Strictly speaking, we should expect to recover the
Euler characteristic of the Hilbert space in the chamber
$R \rightarrow \infty$ (the ``long neck" chamber). However,
one can easily check that, in this chamber, one also has
\casiles\ for the partition function. This is easily seen
by using the wall-crossing formulae derived in \ns\ for the
Donaldson invariants: there is no wall-crossing for the
partition function. This interpretation of \casiles\ as
an Euler characteristic is not easy to check from known
mathematical results, however, as on manifolds with
$b_1 (Y)>0$ the Floer homology has only been constructed
when there is a nontrivial magnetic flux on $Y$, in order
to avoid reducible flat connections \floerb\ (see \ffloer\
for a nice review). In order to interpret our result \casiles,
it is illuminating to compute the partition function when $w_2 (E)$
has the integral lift $\alpha_Y$. We can do the computation
in two different ways. When one uses the lattice reduction
and unfolding, the inclusion of the flux $w_2 (E)= \alpha_Y$
has the following effect: the contribution of the monopole
and the dyon cusps is the same as before, but for the cusps
at infinity, one has to change
\eqn\change{
{1  \over
1-{\rm e} ^{-i  (\tilde S, z)/ h_I }  } \rightarrow {1 \over 2i}
\csc \biggl(  {(\tilde S, z)\over  2 h_I} \biggr).}
After doing this, the contribution from the monopole and the
dyon cancel the contribution from the four semiclassical
regions. Therefore, the $u$-plane integral vanishes.
Alternatively, one can consider the chamber
$R\rightarrow \infty$, where the vanishing theorem
for the $u$-plane integral holds. As there is no
wall-crossing for the partition function, we find
again $Z_u=0$. Therefore,
\eqn\wflux{
Z_{DW}^{w_2=\alpha_Y} (Y \times {\bf S}^1)=2 \Delta_Y''(1).}
We see that the inclusion of a nonzero flux, which gets rid
of the reducibles, kills the $u$-plane
contribution, as expected. The term $-\Delta_Y (1)$ should
be understood as the contribution of the reducible flat
connections on $Y$ to the partition function.

\subsec{A relation to Reidemeister-Milnor torsion}

In the above computations, we have not included any
observable in the generating function. One can try to
include an appropriate 2-observable in such a way that
the Donaldson-Witten function has the structure of a
formal series related to the Meng-Taubes series. When
a 2-observable is included, the SW contribution for
$Y \times {\bf S}^1$ has the structure \monopole\ns
\eqn\swobs{
Z_{SW}({\rm e}^{I(S)})= 2   \sum_{\lambda}
SW(\lambda)\biggl( {\rm e}^{(S,x) + S^2/2} +
{\rm e}^{-i(S,x) - S^2/2}\biggr),}
where we have put $w_2 (E)=0$, the sum is over ${\rm Spin}^c$
structures and $x= 2k \alpha_Y$. If we consider the 2-homology class
\eqn\observ{
S={1\over 2}  t \,  S_2,
}
where $t$ has to be understood as a formal parameter,
we see that the dependence in $t$ has the form
\eqn\formalt{
{\rm e}^{(S,x)}= ({\rm e}^t)^k,}
for the monopole contribution.Therefore, the sum
of SW invariants corresponding to the ${\rm Spin}^c$
structures with $x=2k \alpha_Y$ are the coefficients of
a polynomial in ${\rm e}^t$  (for the monopole contribution)
and in ${\rm e}^{-it}$ (for the dyon contribution),
very much as in \cpseries, and \swobs\ becomes
\eqn\formal{
\sum_{x \in H} \Biggl( \sum_{c|\bar c_1(c)=x}
SW(Y,c) \Biggr) \bigl( ({\rm e}^t) ^k + ({\rm e}^{-it})^k \bigr).}
Notice that the SW invariants considered here
are computed using a small perturbation.
The surprise comes when one computes the
$u$-plane contribution. We have to expand
\eqn\expansion{
{1 \over 1-{\rm e} ^{-i { (\tilde S, z)/ h}}} = {
 1 \over 1-{\rm e} ^{-i { ( S, z) /h}} }  +
{ {\rm e} ^{-i { ( S, z) / h}} \over
\bigl( 1-{\rm e} ^{-i { ( S, z)/ h}} \bigr)^2}
{ i {\sqrt 2} \over 16} {d \tau \over da} \Omega + \dots,}
and only the second term survives after integrating over
the $2$-torus of flat connections. We have to extract
the $q^0$ power of the expansions at the different
cusps. The monopole and dyon cusp contributions are
regular at $q=0$, while the semiclassical cusp gives
a power series in $h_{\infty}(q)$,
where $h_{\infty}$ is a modular form given in \mw.
The final result is
\eqn\surprise{
\eqalign{
Z_{DW} ({\rm e}^{I(S)}) &= 2 { |{\rm Tor}(H_1(Y, \IZ))|
\over  (({\rm e}^t)^{1/2} -({\rm e}^t)^{-1/2})^2} +
2 \sum_{x \in H} \bigl( \sum_{c|\bar c_1(c)=x} SW(Y,c) \bigr)
({\rm e}^t)^k \cr
& + 2 { |{\rm Tor}(H_1(Y, \IZ))|
\over  (({\rm e}^{-it})^{1/2} -({\rm e}^{-it})^{-1/2})^2} +
 2\sum_{x \in H} \bigl( \sum_{c|\bar c_1(c)=x} SW(Y,c) \bigr)
 ({\rm e}^{-it})^k\cr
&- \biggl[ { 2 |{\rm Tor}(H_1(Y, \IZ))|
\over \bigl( \sin (t/4 h_{\infty}) \bigr)^2} \biggr]_{q^0}.\cr} }
This expression is regular when $t=0$, as the
poles cancel between the monopole and dyon cusps. We can
write it in a more compact form using \finwc\ and \mtone:
\eqn\fform{
Z_{DW} ({\rm e}^{I(S)}) = 2 \tau ( Y; {\rm e}^t)
+ 2 \tau  (Y; {\rm e}^{-it}) -
\biggl[ { 2 |{\rm Tor}(H_1(Y, \IZ))| \over
\bigl( \sin (t/4 h_{\infty}) \bigr)^2} \biggr]_{q^0}.}
We see that the infinite series associated
to the stable SW invariants can be
reinterpreted as the $u$-plane contribution from the monopole
or dyon cusps (in the chamber $R \rightarrow 0$) to the
generating function associated to the observable \observ. In addition,
we have found a relation between the Reidemeister-Milnor torsion and
a generating function in Donaldson-Witten theory. It is interesting
to notice that
Donaldson-Witten functions involving
${\rm e}^{t I(S)}$ appear in a natural way in the context of
Fukaya-Floer homology (see for instance \ffloer.)

We should point out that the generalization of the results obtained
here for $b_1(Y)=1$ to the
higher rank case is not an easy task, since the computation of the integral
over the Coulomb branch can not be done using the unfolding technique.

\newsec{On the perils of compactification}

We now return to the key result \comp\ and
investigate its meaning.

\subsec{Review of the relation of three- and four-dimensional
SYM}

In this section we review some results of
Seiberg and Witten \threesw\ and of Rozansky and Witten
\rw.

In \threesw\ Seiberg and Witten studied
the low-energy effective action of ${\cal N}=2$ super
Yang-Mills compactified on $\IR^3 \times {\bf S}^1 $  where
the $S^1$ factor has radius $R$.
They  argued that in the limits $R\rightarrow \infty$
and $R \rightarrow 0$ one recovers the pure 4d and 3d
theories, respectively, and therefore that the two
different limits are connected through an interpolating
theory that depends on $R$. The low-energy description
of the compactified theory is a three-dimensional
$\CN=4$ sigma model
whose target space is a hyperk\"ahler manifold
$\CM_R$. As a complex manifold $\CM_R$ can be
identified with the total space of the elliptic fibration
over the $u$-plane defined by the SW curve.
The metric on $\CM_R$ depends on the
compactification radius $R$ and has not
been determined explicitly for general values of
$R$. In the limit  as $R \rightarrow 0$ there
is a well-defined limit on  the complement
of a section of the elliptic fibration and the
limiting metric turns out to be
the    Atiyah-Hitchin metric.

The derivation of the sigma model
with target $\CM_R$ can be approached in
two ways: One can first work out a low energy
theory in 4 dimensions and then compactify,
or one can compactify and then work out the
quantum corrections.  The first approach is
better at large $R$ and the second is better at
small $R$. We elaborate on this briefly.

The first method uses the   compactification
 of the low-energy SW effective theory
of a $U(1)$ vectormultiplet
\threesw.
In  this point of view we first first work in
four dimensions and go to the infrared.
To write the low energy lagrangian we
{\it choose}  a
duality frame, i.e., we use $SL(2,\IZ)$ to make
 a choice of  weakly coupled
$U(1)$ vectormultiplet $(a(u), A, \lambda) $.
We next carry out dimensional
reduction. Then we use 3-dimensional
dualization to go   from the 3D vector field
$A_\mu$ to a compact
scalar $\sigma$. The result is  the tree level sigma model:
\eqn\swtreelev{
\int_Y    2 \pi R g_{u \bar u} du \wedge * d \bar u
+ {1 \over  8 \pi^2 R \im \tau(u) } \vert d \sigma - \tau d b\vert^2
}
where $0 \leq \sigma ,b \leq 2 \pi$.
\foot{There is one important difference relative to
\threesw. In our case the threefold $Y$ is
compact, with a volume growing like
$\vol(Y) \sim R^{-1}$. } Thus,
the sigma model has
as target the total space of the elliptic fibration over
the $u$-plane. The metric in \swtreelev\
is only an approximation. However, the
underlying complex manifold is exactly
determined.
As a complex manifold the total space over
the $u$-plane is   the surface
\eqn\surface{
z y^2 = x^3 - z x^2 u + z^2 x
}
for   $\bigl( (x:y:z) ; u\bigr) \in \IC P^2 \times \IC$.
As shown in \threesw, after removing
a section of the fibration one may identify
this surface with the Atiyah-Hitchin manifold
in one of its complex structures.
Unfortunately, there are important quantum
corrections and Kaluza-Klein
corrections which are hard to control in
this approach.

Working instead at small $R$
one can make a compactification
 of the the underlying UV $SU(2)$ $\CN=2,d=4$
theory, and then work out the
quantum dynamics. In the
limit $R \rightarrow 0$ we expect that
we can use the dimensional reduction
of the UV theory
 to obtain $SU(2)$ $\CN=4, d=3$ SYM.
In this theory one can study quantum corrections.
Denoting the scalar field vevs in the Cartan subalgebra
by $\vec \phi$, and working at
  large $\vert \vec \phi \vert$ and to one-loop
approximation
one finds a Taub-Nut metric
\seibtd\threesw:
\eqn\tnmetdef{
ds^2 = V^{-1}  (d \sigma + \vec \omega \cdot d \vec \phi)^2 +
V  (d \vec \phi)^2
}
for the target space of the 3D sigma model. Here
$\sigma$ is the dualized photon and, as usual,
$\nabla \times \vec \omega = \nabla V$ \egh.
Moreover in this case the
TN potential here has negative  mass:
\eqn\tnmet{
V = {2 \pi \over  e_3^2} - { 1 \over  2 \pi \vert \vec \phi\vert}
}
Furthermore, studies of 3D instanton effects
reveal the leading $e^{-r}$ corrections corresponding
to the Atiyah-Hitchin metric \khoze.

Motivated by
the non-perturbative results in
supersymmetric gauge theory in three dimensions
of \threesw, Rozansky and Witten constructed a new
topological field theory in three dimensions  \rw.
\foot{A new paper on the subject, with
some relation to the present paper, was recently
posted on the
e-print archives \gthompson.}
The RW theory is
based on the twisting of an ${\cal N}=4$
sigma model with target space a hyperk\"ahler manifold $\CM $.
It has been known for some time that the partition function of the twisted
${\cal N}=4$ Yang-Mills theory in three
dimensions (which is the dimensional reduction of
Donaldson-Witten theory) {\it formally}
computes the Casson invariant,
\topc\aj\bt\btft. In \rw, Rozansky and
Witten used the low-energy description of this theory to
show that this is not just formally true,
but really true in  the case of homology
spheres, while for three manifolds with $b_1 (Y)>0$, the Rozansky-Witten
partition function is precisely Lescop's
extension of the Casson invariant:
\eqn\rwcwl{
Z_{RW}(Y; \CM_0) = \lambda_{CWL}(Y) .
}
More generally, we may use the interpolating
hyperk\"ahler manifold  $\CM_R$ of the SW 3D
sigma model  to obtain:
\eqn\rwcwl{
Z_{RW}(Y; \CM_R) =-{1\over 2}b_\theta(\CM_R) \lambda_{CWL}(Y),
}
where \rw\
\eqn\bthe{
b_{\theta} (\CM_R) =\int_{\CM_R}
{1 \over 8 \pi^2} {\rm Tr} [ \CR \wedge \CR] ,}
and $\CR$ is the curvature two-form
associated to the hyperk\"ahler metric
on $\CM_R$. For $R=0$,   $\CM_R$ is the
Atiyah-Hitchin manifold, and the   integral is $-2$.

\subsec{  Donaldson-Witten $\not=$ Rozansky-Witten }

We have considered the partition function of
Donaldson-Witten theory on the four-manifold
$X= Y \times {\bf  S}^1$ in the limit when the
radius of the circle goes to zero. For manifolds
with $b_1 (Y)=1$ our result does not agree with Lescop's
extension of the Casson invariant, and therefore
does not agree with the Rozansky-Witten partition
function. However, the results are not totally
unrelated and we can be more precise about the
relation between the different quantities.

In general, the Donaldson-Witten function has
the structure
\eqn\usw{
Z_{DW} = Z_u + Z_{SW}
}
but there is no canonical decomposition into
a ``$u$-plane part'' and an ``SW part.''
As we change the metric the relative contributions
change due to SW wall-crossing. However,
when there is no SW wall-crossing, as in the present
case, the decomposition of the Donaldson-Witten function
in terms of SW contributions and the $u$-plane integral
is canonical, since they do not mix.

Moreover, when
we perform the computation of $Z_u$
in a chamber such as
$R \rightarrow 0$ by lattice reduction
and unfolding, the contributions from the different regions
on the $u$-plane are well distinguished: the contribution of
monopole and dyon cusps correspond to finite regions in
the $u$-plane centered around the monopole and dyon
singularities, respectively, while the four
semiclassical cusps correspond to regions that extend to
infinity in the $u$-plane. Thus, given a chamber,  we have
the decomposition
\eqn\udecomp{
Z_u = Z_{u,M} + Z_{u,D} + Z_{u,\infty} .
}
It is important to stress that, in general,
 the decomposition of the $u$-plane
integral into contributions from different cusps is not canonical
and depends on the chamber under consideration (when the integral
is computed using lattice reduction, this decomposition depends on
the chamber through the choice of a lattice vector $z$, as explained in
section 8 of \mw).  However,
in the present case, for {\it both}  $R \rightarrow 0$ and
$R \rightarrow \infty$ we find
\eqn\zeeyou{
Z_{u } = -
\vert {\rm Tor}(H_1(Y, \IZ)) \vert}
because there is neither  Donaldson nor SW
wall-crossing. More surprisingly, we find
for  $R \rightarrow 0,\infty$ the same decomposition:
\foot{
There is a very interesting further subtlety here. If we also
``regularize'' by including a two-observable $I(S)$ as in
\surprise\ then we find that $Z_u(e^{I(S)})$ is different
in the chambers $R \rightarrow 0$ and
$R \rightarrow \infty$. Still when we subsequently
take the limit $S \rightarrow 0$ we obtain the same result:
$\lim_{S \rightarrow 0}  Z_u(e^{I(S)})=   -
\vert {\rm Tor}(H_1(Y, \IZ)) \vert$ in both chambers. Nevertheless,
if we first take $R \rightarrow \infty$ and then let
$S \rightarrow 0$ we find a different decomposition
of the $u$-plane integral:
$Z_{u,\infty} =  -
\vert {\rm Tor}(H_1(Y, \IZ)) \vert$, $Z_{u,M}=0$,
$Z_{u,D}=0$. }
\eqn\udecompii{
\eqalign{
Z_{u,M} &
= Z_{u,D} = -{1 \over  6} \vert {\rm Tor}(H_1(Y, \IZ)) \vert\cr
Z_{u,\infty}
& = -{4 \over  6} \vert {\rm Tor}(H_1(Y, \IZ)) \vert .\cr}
}

Combining these two decompositions we
can write a decomposition of the Donaldson-Witten
function for the chamber $R \rightarrow 0$
\eqn\mdi{
Z_{DW} = Z_M + Z_D + Z_{\infty}.
}
For example, the contribution of
  the monopole cusp
is given by:
\eqn\moncon{
Z_M = 2\biggl( {1 \over 2} \Delta_Y ''(1) - { |{\rm Tor}(H_1(Y, \IZ))|
\over 12} \biggr)= 2 \lambda_{CWL} (Y).
}
Here the first term comes
from the SW invariants at $u=1$, and the second term comes
from the contribution of the monopole cusp in the $u$-plane integral.
The same result holds for $Z_D$. Therefore, this ``truncated" topological
invariant
agrees with the Rozansky-Witten invariant (after including in
the latter the factor of $2$ due to the center of the gauge group),
and therefore with Lescop's extension of the Casson invariant.
In comparing the theories   we therefore have:
\eqn\comparsion{
\eqalign{
Z_{M}(Y \times {\bf S}^1 ) & =2 Z_{RW}(Y) \cr
Z_{D}(Y \times {\bf S}^1 ) & =2 Z_{RW}(Y) \cr
Z_{\infty}(Y \times {\bf S}^1 ) & =
-{4 \over  6} \vert {\rm Tor}(H_1(Y, \IZ)) \vert\   \not\propto  2
Z_{RW}(Y) \cr}
}

If we include the two-observable \observ\  and use
  \surprise\ we find that  $Z_M$ is given by
\eqn\obs{
Z_M ({\rm e}^{I(S)}) =2  \tau (Y;{\rm e}^t).}
It is interesting to notice that the second term in \moncon\
can be interpreted as the $\zeta$-regularization of \obs\ as
$t \rightarrow 0$. The infinite series one obtains in this limit is
precisely the infinite series of wall-crossings \univ:
\eqn\regul{
  \vert {\rm Tor}(H_1(Y, \IZ)) \vert(1+2+ \dots) = \zeta (-1)  |{\rm
Tor}(H_1(Y, \IZ))| = -{ |{\rm Tor}(H_1(Y, \IZ))|
\over 12} .}

A glance at \surprise\ shows that the cusp at infinity
does not contribute anything like the torsion.
It remains to understand more clearly {\it why} there
is a discrepancy between three-dimensional and
four-dimensional theories. Note that this subtlety does
not enter for $b_1(Y)>1$. In this case there is
no $u$-plane contribution and the 4D and 3D theories are
related in the expected way. Therefore, we begin
by revisiting the $u$-plane integral.

\subsec{A closer look at the $u$-plane measure}

Let us now examine more closely the $u$-plane
integral:
\eqn\copy{
\eqalign{
Z_u(Y \times {\bf S}^1)  =
\half \vert {\rm Tor}(H_1(Y, \IZ))  & \vert
\int_{{\Gamma}^0(4) \backslash \CH }
 {dx dy \over y^{1/2}} \cdot
  \cr
 \sum_{n,m\in \IZ } \Biggl\{
\biggl({\rm e}^{\theta}
(n^2 {\rm e}^{-2\theta} - m^2 {\rm e}^{2 \theta})  - {\rm e}^{\theta} {1
\over  2 \pi y}
\biggr) &
\cdot   \exp \biggl[ -\pi y (n^2 {\rm e}^{-2\theta} + m^2 {\rm e}^{2
\theta}) - 2\pi i m x \biggr] \Biggr\} \cr}}
The integral is over a fundamental domain for the
congruence subgroup $\Gamma^0(4)$. We denote
$\tau = x+i y$.
The sum is over  line bundles for the $U(1)$
gauge  theory with
\eqn\latvec{
\lambda= n[S_1] + m [S_2] \qquad n,m\in \IZ .}
Recall that the metric defines the period point $*\omega= \omega$ with
\eqn\per{
\omega = {1 \over {\sqrt 2}}({\rm e}^{\theta} [S_1] + {\rm e}^{-\theta}
[S_2]),}
 The first term in the sum in
\copy\ comes from bringing down the
term $\sim { d \tau \over  da} F \psi \psi $ in the action and soaking up
fermion $\psi$-zeromodes associated with $b_1(X) = 2$.
The second term in the sum in
\copy\ is a contact term.

Referring to the definitions of the classes
$[S_1],[S_2]$ and to
\volume\helpful\ we see that the limit of shrinking
Kaluza-Klein circle  $R  \sim e^{- \theta} \rightarrow 0$ corresponds to
$\theta \rightarrow \infty$.
Let us now consider the behavior of
$Z_u$ in this limit.

The first thing to note is
 that the terms in the integrand with $m=0$
actually blow up in this limit! The reason for this is that
such line bundles have ``instanton'' connections
with    no dependence in the
Kaluza-Klein circle direction $\varphi$.
However, since the overall
volume is fixed, the volume of $Y$ goes like
$e^{+\theta}\rightarrow + \infty$. Thus, new zeromodes,
related to the decompactification of $Y$ develop,
causing a term by term divergence in the
integrand of \copy.
Interestingly, there is in fact a {\it cancellation} between
the positive contribution of the fermion zeromode
term and the negative contribution of the contact
term. This can be seen mathematically as follows.

First, note that the two terms in the
sum in \copy\  combine as a total derivative in
$\theta$:
\eqn\totalderv{
{e^{2 \theta}  \over 2 \pi y } {d \over d\theta}
\Biggl[
e^{-\theta} \sum_{n,m} \exp \biggl\{
 -\pi y (n^2 {\rm e}^{-2\theta} + m^2 {\rm e}^{2 \theta}) - 2\pi i m x \biggr\}
\Biggr] .
}
Now we see that the divergence from the
sum on $n$ (at $m=0$) can be offset by the vanishing
of $e^{-\theta}$. To see which dominates we use
the Poisson summation formula to write:
\eqn\poi{
\sum_{n,m}
\exp \biggl[ -\pi y (n^2 {\rm e}^{-2\theta} + m^2 {\rm e}^{2 \theta}) -
2\pi i m x \biggr] =
{ {\rm e}^{\theta} \over y^{1/2} } \sum_{\hat n, m} \exp \biggl[ -{\pi
\over y} |\hat n + m \tau|^2  {\rm e}^{2\theta} \biggr].}

Combining the two terms one sees that \copy\ becomes
\eqn\lastint{Z_u =
- {1 \over 2} \vert {\rm Tor}(H_1(Y, \IZ))
  \vert \int_{{\Gamma}^0(4) \backslash \CH }
   {dx dy \over y^{3}} {\rm e}^{4 \theta} \sum_{\hat n, m} |\hat n + m\tau|^2
    \exp \biggl[ -{\pi \over y} |\hat n + m \tau|^2  {\rm e}^{2\theta}
\biggr] .}

We can learn two things from \lastint. First, we now
note that not only have the divergences of the $m=0$
terms cancelled, but the remaining integrand actually
 {\it vanishes}
exponentially fast as $\theta \rightarrow + \infty$:
\eqn\limintgnd{
\lim_{\theta \rightarrow + \infty}
{\rm e}^{4 \theta} \sum_{\hat n, m} |\hat n + m\tau|^2
\exp \biggl[ -{\pi \over y} |\hat n + m \tau|^2  {\rm e}^{2\theta} \biggr] =0 .
}

The second thing we learn from \lastint\ is that
the measure is in fact $SL(2,\IZ)$ invariant.
This is a surprise since in general the $u$-plane
measure
is only   ${\Gamma}^0(4)$ invariant.
Thus, each of the six copies of the
fundamental domain $\CF$ of $SL(2,\IZ)$
contribute equally to $Z_u$. Moreover,
the integrand is in the standard form for
which one can apply the unfolding technique.
Combining these two observations we get:
\eqn\unfold{
Z_u = - 3 \vert {\rm Tor}(H_1(Y, \IZ))
\vert \int _0^{\infty}{ dy \over y^{3}} {\rm e}^{4 \theta}
 \sum_{\hat n} \hat n^2 \exp \biggl[ -\pi  \hat n^2  {e^{2 \theta} \over y}
 \biggr] .}

We can some further insight into the nature of
the measure from the expression \unfold.
Note first that we can explicitly eliminate all
$\theta$-dependence by a change of variables to
\eqn\chgvrbl{
\xi \equiv {e^{2 \theta } \over  y} =  {1 \over  2 R^2 y} .
}
Thus the integral \unfold\ is in fact $R$-independent and
simply given by $- \vert {\rm Tor}(H_1(Y, \IZ)) \vert $.
As we have observed, and as is even more
obvious from \unfold, at {\it fixed} value of
$y$, as  $R \rightarrow 0 $ the integrand vanishes.
On the other hand the integral is $R$-independent
and nonzero. Thus the
integrand is becoming delta function supported.
We can see this rather explicitly by letting
$w=1/y$ and noting that:
\eqn\limdelts{
\eqalign{
 \lim_{R \rightarrow 0}  \biggl(  \sum_{\hat n \in \IZ} \hat n^2 {w \over
R^4} e^{-\pi \hat n^2 w/R^2}
\biggr) dw & =
 \sum_{\hat n \not=0 } \lim_{R \rightarrow 0 }
\hat n^2 {w \over  R^4} e^{-\pi \hat n^2 w/R^2}  dw \cr
& =
 \sum_{\hat n \not=0 } {1 \over  \pi^2 \hat n^2} \delta(w)  dw\cr
& = {1 \over  3} \delta(w) dw \cr}
}
Thus, as $R \rightarrow 0$, the measure in each cusp
region is becoming $\delta$-function supported at
$y=\infty$.\foot{This is similar to the source of the holomorphic
anomaly in certain one-loop expressions in string
theory \bcov.}
  Now, this discussion can be carried out in each of
the three cusp regions of the fundamental
domain of ${\Gamma}^0(4)$. In each cusp region
we have a different $q$ expansion $q = e^{2 \pi i \tau} $,
$\tau = x + i y$. Denoting the relevant $q$-parameters
by $q_M, q_D, q_{\infty}$  the $R \rightarrow 0$ limit
of the $u$-plane  measure is
\eqn\arrzeromes{
-\vert {\rm Tor}(H_1(Y, \IZ)) \vert \bigg\{
{1 \over  6} \delta^{(2)}(q_M) d^2 q_M +
{1 \over  6} \delta^{(2)}(q_D) d^2 q_D +
{4 \over  6} \delta^{(2)}(q_\infty) d^2 q_\infty
\biggr\} .
}

\subsec{An interpretation of the result}

Given the facts reviewed in section 6.1, the
discrepancy between
$Z_{DW}(Y \times {\bf S}^1)$ and $Z_{RW}(Y)$
is somewhat surprising. In this section we
will discuss some of the physics behind this
discrepancy and suggest an interpretation
of the result. We thank N. Seiberg and E. Witten
for important remarks that helped us to this
picture.

Let us first dispose of a red-herring. Nonintegral values of
the Witten index are often associated with the
presence of
noncompact field spaces, and the mishandling of
 a ``bulk'' or a ``boundary'' contribution.
We stress that this is {\it not} what is going on
here since $Z_{DW}(Y \times {\bf S}^1)$ has no wall crossing.

Our interpretation of \comparsion\  is that the
Donaldson-Witten theory on
$X=Y \times {\bf S}^1$ for small $R$ is simultaneously a
three-dimensional and a four-dimensional theory.
By this we mean the following:
We must integrate over moduli space to get the physical
partition function. There is very different
physics in the different regimes of moduli space.
Some of it is three-dimensional and some of
it is four-dimensional.

For small $R$ the measure for the cusp at $\infty$ is
concentrated in the region
\eqn\infreg{
\im \tau_{\infty}(u) \gsim 1/R^2
}
where $\tau_{\infty}$ is the $\tau$ parameter
selected by the semiclassical cusp. Because
of asymptotic freedom,  at small $R$ we can use the semiclassical one-loop
answer and the
measure is concentrated in
the region
\eqn\inftyregii{
\log \vert u \vert \gsim  {\pi \over  2 R^2}
}
In this region of the $u$-plane physics is
effectively four-dimensional. The infrared
4-dimensional SW description becomes applicable
at length scales
\eqn\stoprun{
\ell \sim {1 \over  \sqrt{u}} \sim \exp[ - { \pi \over  4 R^2} ]
\ll R
}
At such length scales the compactification on ${\bf S}^1_R$ is
completely irrelevant. Because of asymptotic freedom
this becomes better and better as $R \rightarrow 0$.

Let us now consider the monopole cusp. The $u$-plane
measure is concentrated in the region
\eqn\infreg{
\im \tau_{M}(u) \gsim 1/R^2
}
where $\tau_M = -1/\tau_{\infty}$ defines the weak-coupling
frame near the monopole cusp $u=1$.   In particular
$\tau_M(u) \cong { 1\over  i \pi} \log (u-1) $
so the relevant region of the $u$-plane is:
\eqn\monreg{
\vert u-1 \vert
{\ \lower-1.2pt\vbox{\hbox{\rlap{$<$}\lower5pt\vbox{\hbox{$\sim$}}}}\ } e^{
- \pi /R^2}
}
consequently the monopoles are very light.

However, the effective theory of monopole
hypermultiplets and dual $U(1)$ vectormultiplets is
IR free and UV unstable - it is not defined as a
four-dimensional theory at distance scales $\ell \ll R$.
Indeed, the infrared SW description is only applicable
at length scales
\eqn\stoprunii{
\ell \gsim e^{+ \pi/R^2} \gg R
}
For this region of moduli space we must first
compactify and then solve for the dynamics.

\subsec{Comments on one-loop corrections}

When one combines standard one-loop expressions
with some of the above remarks one can be lead to
paradoxes which have troubled the authors
of this paper not a little. In this section we
mention some of these confusions, and suggest a
resolution.

In classical dimensional reduction
the gauge couplings
$e_3^2$ and $g_4^2$ in three and four
dimensions, respectively,
  are related by $g_4^2 = e_3^2 R$.

In 4D gauge theory, when integrating out massive
charged vectormultiplets and hypermultiplets
of mass $m_i$ and charge $Q_i$ in
a weakly coupled theory the  threshold
correction relating the coupling $g_{4,UV}^2$ of the
underlying theory and $g_{4,IR}^2$ of the low
energy effective theory is:
\eqn\fdren{
{ 8 \pi \over  g_{4,IR}^2} =
{ 8 \pi \over  g_{4,UV}^2} - 16 \pi
\sum_i (-1)^{\epsilon_i} Q_i^2 \int {d^4 p \over  (2 \pi)^4}
{1 \over  (p^2 + m_i^2 )^2}
}
where $\epsilon_i =  0$ for VM's and $\epsilon_i =1$ for HM's.
The integral in \fdren\ is log divergent and a regularization
\eqn\regular{
{1 \over  (p^2 + m_i^2)^2 } \rightarrow { \Lambda^{2 \alpha  -4 } \over
(p^2 + m_i^2)^\alpha}
}
with $\alpha = 2 + \epsilon$, $\epsilon \rightarrow 0^+$
is understood here and below.

When we compactify $\IR^4 \rightarrow \IR^3 \times {\bf S}^1$
the integral in  \fdren\ becomes
\eqn\intren{
{1 \over  R}\sum_{n=-\infty}^{\infty}
\int {d^3 \vec p \over  (2 \pi)^3}
{1 \over  (\vec p^2 + (A_4 + n/R)^2 + m^2 )^2}
}
where $A_4$ is a background Wilson loop.
The expression \intren\ interpolates nicely between
the renormalizations in 3D and 4D.
\foot{We elaborate
here on remarks in \sh\ssh. }
Indeed, performing the integral on
$\vec p$ we get:
\eqn\vcpint{
{\pi^{3/2} \over  R} {\Gamma(\alpha-3/2) \over  \Gamma(\alpha)}
\sum_{n=-\infty}^{\infty}
{ \Lambda^{2 \alpha  -4 }  \over  ((A_4 + n/R)^2 + m^2)^{\alpha-3/2} } .
}
At small values of $R$ we have
\eqn\vcpintii{
{1 \over  \epsilon} + { \pi^2 \over  R} { 1 \over  \sqrt{A_4^2 + m^2}}
+ F(RA_4 , R m)
}
where $F(x,y)$ is an analytic series vanishing
as $x,y \rightarrow 0$.

On the other hand, at large values of $R$ we find
for the same integral:
\eqn\largearr{
{\pi^2 \over  \epsilon} - \pi^2 \log  {m^2 \over  \Lambda^2}
+ 2 \pi^2 \sum_{n\not=0} e^{ 2 \pi i n (A_4 R)}
K_0(2 \pi \vert n \vert m R)
}

The physics of \largearr\ is
 clear: The log term is the 4d 1-loop
effect of integrating out heavy charged
particles in the
low energy effective abelian theory.
 The Bessel functions from the nonzero modes
decay as $\sim
{1 \over  2 \sqrt{\vert n \vert m R}} e^{-2 \pi \vert n \vert m R} $
and can be understood, from the 3D perspective, as
instantons from particles of mass $m$ running in the
${\bf S}^1$ loop. As pointed out in
\threesw, such  quantum corrections are expected
to renormalize the metric to  a smooth hyperk\"ahler metric on the moduli
space. The expression
\vcpintii\ can then be understood as a modification
\eqn\tnmodfy{
{8 \pi \over  e_{3,IR}^2 } = {8 \pi \over  e_{3,UV}^2 } -
 16 \pi
\sum_i (-1)^{\epsilon_i} Q_i^2
{ \pi^2 \over  \sqrt{A_4^2 + m^2}}
}
Identifying
$   A_4^2 + m^2$ with  $\vec \phi^2$
we reproduce the 3D 1-loop result. Dualization
of the photon then leads to a Taub-NUT metric
with
\eqn\tnmassdef{
V \propto 1 + { M_{TN} \over  r}
}
Here $M_{TN}$ is the ``Taub-NUT mass,''
and $r$ is the radial Euclidean distance in the standard
representation of the TN space as a circle fibration
over $\IR^3$ \egh.

Comparing \tnmodfy\ and \tnmassdef\ we see that
there is a direct connection between the sign of
the TN mass and the sign of the coefficient
of the 4D $\beta$-function. A {\it negative}
mass $M_{TN}$
in 3D corresponds to an asymptotically free
beta function in 4D. This leads to an apparent
paradox: The SW 3D target space is the
Atiyah-Hitchin manifold
$\CM_0$ as $R \rightarrow 0$.  The latter
is well-known to be approximately a negative
mass TN space for $r \rightarrow \infty$. How
is this consistent with the concentration of the $u$-plane
measure at $u=\pm 1$ for $R \rightarrow 0$ ?
Indeed, when one examines the detailed map between
$u$-plane coordinates \surface\ and coordinates
$\sigma, \vec \phi$ on the Atiyah-Hitchin manifold
(such as those used in \tnmetdef)  one finds a
complicated relation between ``regions at infinity.''
In particular, regions of large
$\vert \vec \phi \vert$ can sit over finite points on
the $u$-plane. Since the effective theory
near $u=\pm 1$ is IR free one might expect to
see a {\it positive} mass TN metric. How is this
possible !?

The way out of this confusion is to note that the
4D one-loop analysis in this regime is not very
meaningful. In particular, the monopoles are very
light. From \monreg\ we see that in the relevant
portion of the $u$-plane they have a mass of
order $\vert a_D(u) \vert \lsim e^{-\pi/R^2}$,
and hence expansions such as \largearr\
do not converge.

Clearly, there is much more to understand here,
but we leave it at that, for the moment.

\subsec{Possible implications for   string duality}

The low energy dynamics of D-branes and M-branes
gives a novel and powerful approach to investigating
supersymmetric Yang-Mills theory
\giveonkutasov. Conversely, results on supersymmetric
gauge theory will probably teach us important things
about branes. Here we make a preliminary remark on
a possible implication of the present results for
brane physics.

In the gauge-theory-from-D/M-brane framework, the naive
equivalence of Donaldson-Witten/Seiberg-Witten theory on
$Y \times {\bf S}^1$ to Rozansky-Witten/Seiberg-Witten theory on
$Y$ can be easily proved using standard string
dualities. To do this one begins with the description
of $d=4, \CN=2$ theory as the low energy theory of
an $M5$ brane with two dimensions wrapped on
the Seiberg-Witten curve \wittfdsol.
In the IIA limit the configuration is described by
parallel solitonic 5branes connected by $D4$-branes
as in the Hanany-Witten setup \hanwit.
If the solitonic 5branes wrap $Y \times {\bf S}^1$ we can
apply $T$-duality along the ${\bf S}^1$,
and then $S$-duality to obtain an
effective 3D theory whose low energy dynamics is
described by monopole moduli spaces, such as
$\CM_0$. Our computation shows that, at least for
some quantities, like the partition function with
supersymmetry preserving boundary conditions,
the application of duality should
be applied with care.

\newsec{Conclusions}

In this paper we investigated
the Donaldson-Witten partition function
$Z_{DW} $ on $Y \times {\bf S}^1$
for $b_1(Y)>0$. We have found some interesting relations
to the torsion of $Y$, reinterpreting the result of
Meng and Taubes from the physical point of view,
and gaining some information on Floer homology.

Some very interesting questions remain open.
One important circle of questions is related to
the rational homology spheres with $b_1(Y)=0$.
These present new challenges since, in evaluating
$Z_{DW}$ one must integrate over the $u$-plane
with a density involving one-loop determinants.
Ironically, the actual $u$-plane integral turns out
to be trivial and is just the volume of the
fundamental domain of $\Gamma^0(4)$.
However, this is more than compensated by
the subtleties of the required one-loop graphs.
We defer a discussion of this
subject to another occasion.

We have also discussed some interesting
subtleties in dimensional compactification of
SYM. It would be nice to understand more
deeply than we have done here the origin of
the discrepancy between $Z_{DW}(Y \times {\bf S}^1)$
and $Z_{RW}(Y)$ for manifolds of
$b_1(Y)=1$. A good understanding of the
hyperk\"ahler metric on $\CM_R$ and the relation
between regions at infinity in the $u$-plane and
Atiyah-Hitchin descriptions would be very
helpful.

Finally,  our discussion
has some potential applications in string duality,
as mentioned in section 6.4.

\bigskip
\centerline{\bf Acknowledgements}\nobreak
\bigskip

We would like to thank P. Kronheimer, T. Li, M. Marcolli,
G. Meng, V. Mu\~noz, N. Seiberg and B.L. Wang
for very useful discussions and correspondence. We are specially
indebted to E. Witten for many useful discussions and for his
observations on a preliminary version of this paper.
This work  is supported by
DOE grant DE-FG02-92ER40704.

\listrefs

\bye